\documentclass[amsmath,prb,twocolumn,cite]{revtex4}
\usepackage{verbatim}
\usepackage{graphicx}
\usepackage{amsthm}
\makeatletter

\providecommand{\linkable}[1]{#1}

\makeatother

\begin{document}
\newcommand{\h}{H}

\newcommand{\e}{\mathcal{E}}

\newcommand{\x}{\mathbf{x}}

\newcommand{\vk}{{\mathbf{k}}}

\newcommand{\dx}{\Delta\x}

\newcommand{\fe}{\mathcal{F}_{\text{elast.}}}

\newcommand{\fs}{\mathcal{F}_{\text{surf.}}}

\newcommand{\fw}{\mathcal{F}_{\text{wet.}}}

\providecommand{\cs}[1][SOMETHING]{[CITE:#1]\marginpar{C} }

\providecommand{\linkable}{}

\providecommand{\bf}{\boldsymbol}

\providecommand{\degree}{^\circ}

\providecommand{\n}{{n_\text{cor}}}

\title{Effects of elastic heterogeneity and anisotropy on the morphology
of self-assembled epitaxial quantum dots}

\author{Chandan Kumar and Lawrence Friedman}

\affiliation{Department of Engineering Science and Mechanics, Pennsylvania State
University, 212 Earth and Eng. Sci. Bldg., University Park, PA 16802,
USA.}

\begin{abstract}
\noindent Epitaxial self-assembled quantum dots (SAQDs) are of both technological and fundamental interest, but their reliable manufacture still presents a technical challenge.  To better understand the formation, morphology and ordering of epitaxial self-assembled quantum dots (SAQDs), it is essential to have an accurate model that can aid further experiments and predict the trends in SAQD formation. SAQDs form because of the destabilizing effect of elastic mismatch strain, but most analytic models and some numerical models of SAQD formation either assume an elastically homogeneous anisotropic film-substrate system or assume an elastically heterogeneous isotropic system. In this work, we perform the full film-substrate elastic calculation.  Then we incorporate the elasticity calculation into a stochastic linear growth model.   We find that using homogeneous elasticity can cause errors in the elastic energy density as large as $26\%$, and for typical modeling parameters lead to errors of about $11\%$ in the estimated value of average dot spacing. We also quantify the effect of elastic heterogeneity on the order estimates of SAQDs and confirm previous finding on the possibility of order enhancement by growing a film near the critical film height.
\end{abstract}

\maketitle

Copyright (2008) American Institute of Physics. This article may be downloaded for personal use only. Any other use requires prior permission of the author and the American Institute of Physics. The following article has been submitted to Journal of Applied Physics. After it is published, it will be found at (http://jap.aip.org).

\section{Introduction}

\label{sec1} 

Self-assembled quantum dots (SAQDs) function as artificial atoms embedded
in a semiconductor matrix.~\cite{Bimberg99} As such, they are useful
for a range of electronic and optoelectronic applications.~\cite{Bayer2000,Akiyama:fiberoptics,Viktorov:Laser,Kane:QantumComputer,Elzerman:QuantumComputer,Tanner:QuantumComputer,Li:1996fk,Li:1997lr,Bimberg99,Pchelyakov2000,Grundmann2000,Petroff2001,Liu2001,Heinrichsdorff1997,Bimberg2002,Friesen2003,Cheng2003,Krebs2003,Sakaki2003}
For this reason, there has been a great deal of modeling work on their
dynamic formation process.~\cite{Spencer:1991we,Spencer:1993vt,Tekalign:2004jh,Tekalign:2007,Obayashi:1998fk,Ross:1998fk,Ozkan:1999gf,Gao:1999ve,Holy:1999th,Ortiz:1999ys,Zhang:2003tg,Zhang:2001cr,Meixner:2003lz,Liu:2003kx,Liu:2003qi,Golovin:2003ms,FriedmanStochastic,FriedmanJNP07,Kumar:2007fk,Ramasubramaniam:2005oq,Ramasubramaniam:2005vn,Niu:2006fk}
SAQDs are fabricated by depositing a semiconductor film on a lattice
mismatched substrate with a smaller band gap, the most well-known
examples being Ge$_{x}$Si$_{1-x}$ deposited on Si and In$_{x}$Ga$_{1-x}$As
deposited on GaAs. A good quantitative model of the SAQD formation
can help in aiding understanding of the SAQD formation process and
enable a sophisticated quantitative interpretation of experimental
data, but more importantly, it can help move modeling from a descriptive
mode to a predictive mode that could be used for process design optimization
to aid in tasks such as the formation of new structures, control of
morphology and enhancing order and reproducibility. Here, we improve
upon previous spectral models of SAQD formation~\cite{Spencer:1991we,Spencer:1993vt,Tekalign:2004jh,Tekalign:2007,Obayashi:1998fk,Ozkan:1999gf,Golovin:2003ms,FriedmanStochastic,FriedmanJNP07,Ramasubramaniam:2005oq,Ramasubramaniam:2005vn} by incorporating elastic
anisotropy and elastic heterogeneity simultaneously. We also estimate
the errors introduced by neglecting these effects.

\begin{figure}
\centering
\includegraphics[width=8.5cm,keepaspectratio]{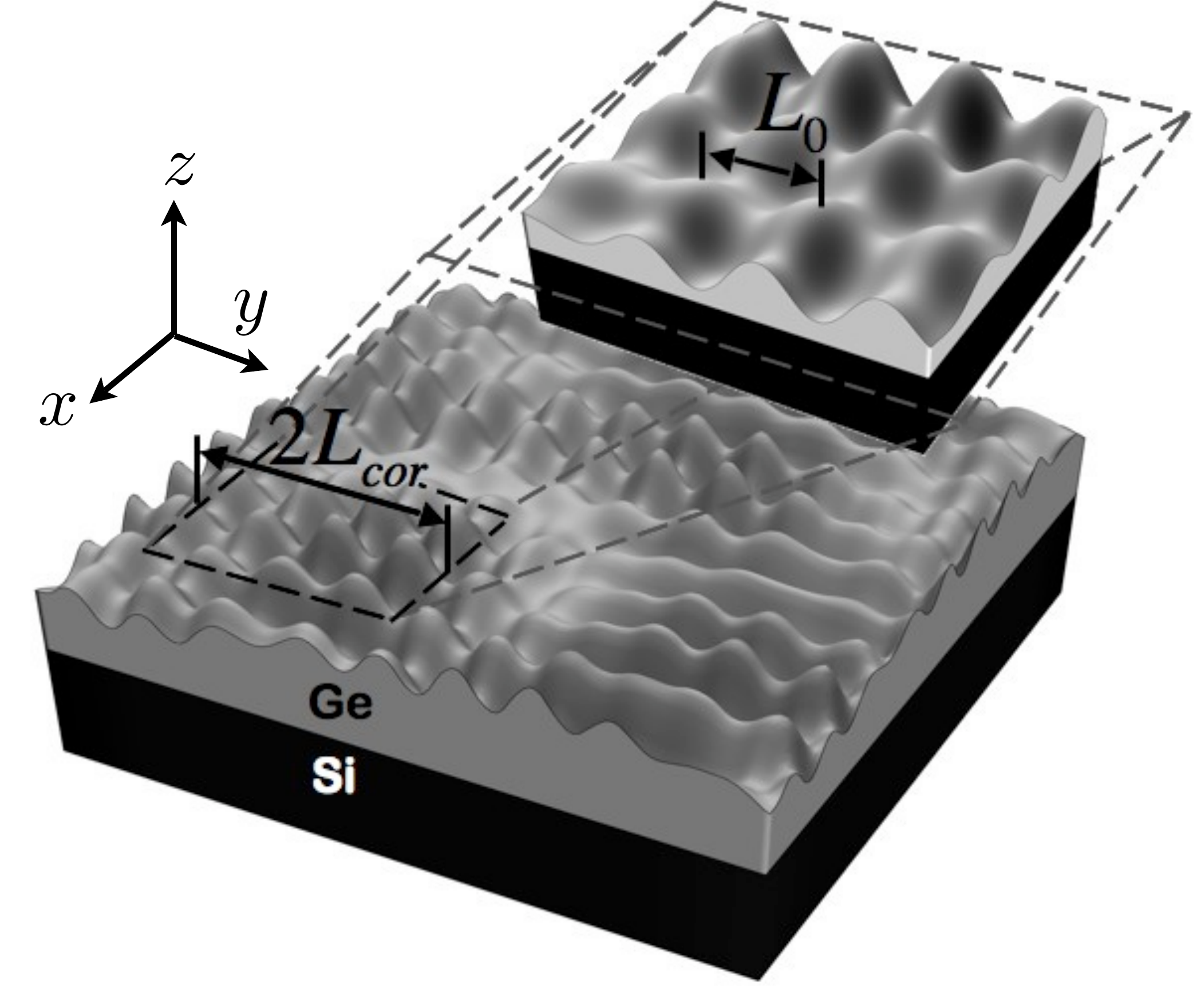}

\caption{Initial formation of Ge/Si SAQDs (Sec.~\ref{sec: linearization det model}) with exaggerated height fluctuations for clarity. $L_0$ is mean dot spacing.  $2 L_{cor}$ is the length over which SAQDs appear periodic. (Sec.~\ref{sec:order analysis}).\label{fig:3Dfigurefilmheight}}
\end{figure}

Many reports in the literature make approximations such as assuming elastic
isotropy,~\cite{Spencer:1991we,Spencer:1993ve,Golovin:2003ms,Kumar:2007fk}
elastic homogeneity of the film-substrate system~\cite{Obayashi:1998fk,Ozkan:1999gf,FriedmanStochastic,FriedmanJNP07,Ramasubramaniam:2005oq,Ramasubramaniam:2005vn}
or making a thin film approximation.~\cite{Tekalign:2004jh,Tekalign:2007}
Here, we present a linear stochastic model of SAQD formation that
incorporates anisotropic elasticity and the elastic heterogeneity of
the film-substrate system. We investigate the SAQD spacing (a mean
property) as well as the order of SAQD arrays that is determined by
the fluctuations in the spacing and alignments (Fig.~\ref{fig:3Dfigurefilmheight}). Furthermore, we investigate
the amount of error that previous approximations make (Table~\ref{tab:compare}). 

While there
are other aspects of SAQD modeling that can be improved or incorporated, the presented
work is an indispensable step in moving toward a more quantitatively
accurate SAQD formation model. The elasticity portion of the calculation
applies generally to different material systems, but other parts of
the calculation such as surface energies and diffusional dynamics
are specific to group IV elements that have four-fold symmetric SAQD
formation dynamics such as Ge$_{x}$Si$_{1-x}$/Si, ~\cite{Friedman:fk,FriedmanJNP07,FriedmanStochastic}
and not to II-VI systems or III-V systems such as In$_{x}$Ga$_{1-x}$As/GaAs.

SAQDs result from a transition from two-dimensional film growth to three-dimensional
growth in strained epitaxial films. When
a flat strained film is perturbed by a film height undulation, elastic
energy is released. When the released energy is greater than the cost
in surface and wetting energy, the perturbation grows. This phenomenon
is known as the ATG (Asaro-Tiller-Grinfeld) instability.~\cite{Asaro-Tiller,Grinfeld}
Eventually the surface perturbations mature into 3D quantum dots.
At a later stage the dots ripen,~\cite{Liu:2003kx,Ross:1998fk} although
theoretically, they might form a uniform array under some circumstances.~\cite{Wang:2004dd,Golovin:2003ms,Ross:1998fk,Ortiz:1999ys,Rastelli:2005kl,Rastelli:2006wd,Tu:2004tg}
The interplay of the elastic energy, the surface energy and
the wetting energy determines the energy landscape that drives SAQD
formation, and the spectral modeling method used here yields a very transparent
description of this interplay. 

The spectral model can be used to define and estimate parameters characterizing
SAQD formation such as the characteristic length and
time scales, the mean SAQD spacing, the alignment of SAQDs in an array
as well as the critical film height for SAQD formation.\cite{Spencer:1991we,Spencer:1993ve,Spencer:1993vt,Ozkan:1999gf,Friedman:fk,FriedmanJNP07}
In the stochastic form, spectral modeling can also elucidate order
and reproducibility of SAQD arrays.~\cite{FriedmanStochastic} In
higher order versions of spectral models known as a multiscale-multitime
analyses, they can even elucidate longer term evolution of SAQDs.~\cite{Golovin:2003ms}
The clearly defined parameters from these models also inform finite
element based models~\cite{Zhang:2003tg} and provide bench-marking
for their performance.

Elasticity is the most well understood influence on SAQD
formation. As such, making fewer approximations about elasticity will
help investigations into other influences on SAQD formation that are
more difficult to understand. For example, wetting energy is barely
understood~\cite{suo98,Zhang:2003tg,Beck:2004yq}, and surface energy
is generally treated as a constant, even though it almost certainly
has strain and temperature dependence. It is also controversial as
to whether surfaces should be treated as facets or not.~\cite{Shenoy:2002lr,Tersoff:2002qf}

While the importance of getting accurate estimates for a quantity
as basic as the mean dot spacing is self-evident, the significance
of SAQD order deserves further discussion. The ordering of SAQDs
has been matter of concern in fostering the development of quantum
dot based devices.~\cite{Hong2006} There are two types of order,
spatial and size. Spatial order is concerned with the uniformity of
the spacings between the SAQDs and size order is concerned with the
uniformity in the size of the SAQDs. The size and spacings of these self-assembled quantum dots
are related, as dot volume is limited by the locally available material.
Understanding what factors affect the order of SAQDs can guide experiments and simulation efforts and help in interpreting experimental and simulation results. Our enhanced elasticity calculation improves recent
models of SAQD order.~\cite{Friedman:fk,FriedmanJNP07,FriedmanStochastic}  We
defer pattern fidelity in directed self-assembly to later work although
some initial results have been previously reported.~\cite{Zhao:2006qy,Kumar:2007fk}

Previous modeling work has focused on how elastic anisotropy and elastic
heterogeneity affect SAQD formation,~\cite{FriedmanJNP07,FriedmanStochastic,Obayashi:1998fk,Spencer:1993ve,Tekalign:2004jh,Golovin:2003ms,Liu:2003ig,Ozkan:1999gf}
but the two influences have been treated separately. The effect of
elastic anisotropy has been studied in great detail. In Ref. \cite{Obayashi:1998fk}
it was shown that for heteroepitaxial system such as $\text{Si}_{1-x}\text{Ge}_{x}/\text{Si}$,
the surface undulations are likely to grow in the $\langle100\rangle$
directions. It was also shown that for anisotropic materials the growth
rate of the amplitude of the surface fluctuations is maximum when
the wavelength is $4/3$ the cutoff wavelength, similar to the isotropic approximation. However, in the presence of
a strong wetting effect, this ratio increases to $2$. In the absence
of misfit dislocations, the islands are aligned in the $\langle100\rangle$
directions. However, experiments reveal that for films with thickness
greater than the critical thickness for dislocation formation, in
the later stages of island formation, the islands align along the
$\langle110\rangle$ directions due to formation of misfit dislocations.~\cite{Ozkan:1999gf} In Ref. \cite{Liu:2003ig} a numerical investigation
was carried out to study the effect of anisotropic strength on the
formation, alignment and average island spacing. More recent analytic studies on SAQD order~\cite{Friedman:fk,FriedmanJNP07,FriedmanStochastic} complement these numerical studies.
The effect of elastic heterogeneity, however, has received more limited
attention. In Ref. \cite{Spencer:1993ve} a linear stability analysis
was performed that incorporated the elastic stiffness for both film
and the substrate. One major conclusion was
that elastically stiff substrate has stabilizing effects on the film that
diminishes with increasing film thickness. In
Ref. \cite{Tekalign:2004jh} a nonlinear evolution equation was derived
using a thin-film approximation. However, Refs.~\cite{Spencer:1993ve}
and~\cite{Tekalign:2004jh} approximate elasticity as isotropic.

\begin{figure}[!h]
\centering
\includegraphics[width=8.5cm,keepaspectratio]{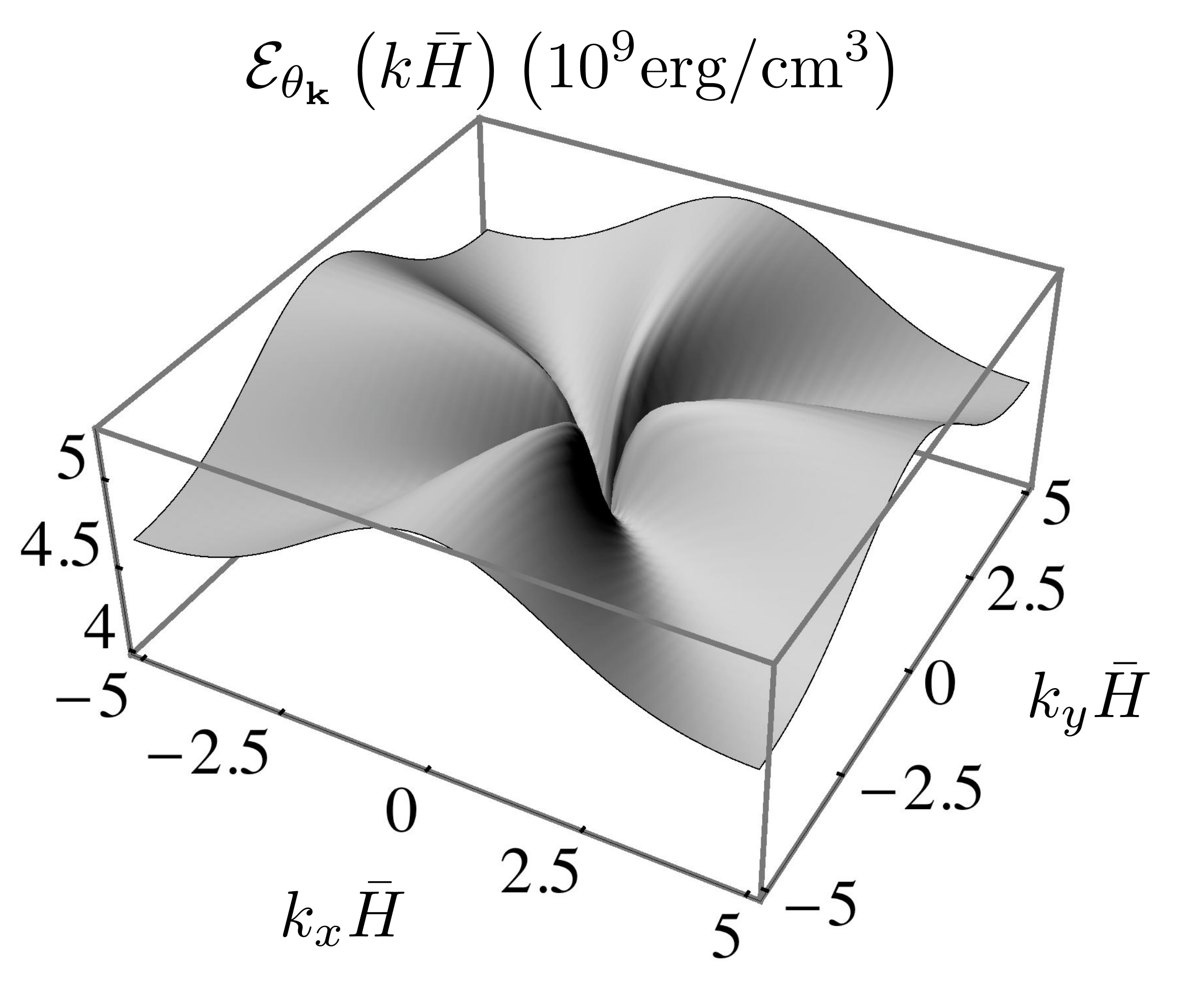}
 \caption{The elastic energy density prefactor as a function of the dimensionless vector $\vk\bar{\h}$ for  Ge/Si at $600\,\text{K}$.\label{fig:3D prefactor}}
\end{figure}

Here, we treat elastic effects without approximations regarding
isotropy, homogeneity or film thickness. We find various parameters that can be derived and estimated from spectral SAQD growth models, and we compare them with the results of the other more approximate models (Table~\ref{tab:compare}).  These parameters are $\e$, the elastic energy density coefficient (Figs.~\ref{fig:3D prefactor} and~\ref{fig:2D prefactor}) $L_E$, the perturbation wavelength that is the most energetically unstable (Fig.~\ref{fig:L_E}), $L_0$, the perturbation wavelength that is kinetically most unstable and gives the mean dot spacing (Fig.~\ref{k_0Sigmak}), and $\n$ (Fig.~\ref{fig:stochastic no of dots}), the number of dots in a row whose positions are well correlated.  Each of these values is compared with the predictions of more approximate models, namely, the elastically \emph{anisotropic homogeneous} approximation, the elastically \emph{anisotropic thin-film} approximation, the elastically \emph{isotropic heterogeneous} approximation, the elastically \emph{isotropic homogeneous} approximation and the elastically \emph{isotropic thin-film} approximation presented recently.~\cite{Tekalign:2007}  For the order analysis, $\n$, comparisons are only made with the elastically \emph{anisotropic} models as elastically \emph{isotropic} models are not suitable for order predictions of periodic arrays.~\cite{Friedman:fk}  Also, all of the reported estimates depend on the average film height ($\bar{\h}$); thus for each comparison we present a calculation corresponding to a typical average film height of $\bar{\h}=4.25\text{ ML} = 1.2\text{ nm}$ in Table~\ref{tab:compare} with some additional values displayed in Figs.~\ref{fig:2D prefactor}, ~\ref{fig:L_E} and~\ref{k_0Sigmak}--\ref{fig:stochastic no of dots}. It is worth noting that all of these approximations correspond to various limits of our elasticity calculation.  For example, the \emph{homogeneous anisotropic} approximation is identical to the limit as the average film height ($\bar{\h}$) becomes large.  The \emph{anisotropic thin-film} approximation corresponds to the limit as $\bar{\h}\rightarrow 0$, and the various isotropic approximations can be obtained by using an isotropic elastic stiffness tensor by, for example, taking the Voigt or Reuss average of the actual elastic moduli. Finally, we give in-depth analysis throughout only for the anisotropic models.

We model the SAQD growth process with a 
stochastic surface diffusion model. We perform a linear analysis
of the dynamics of the film evolution, which corresponds to small height
fluctuations. Although such an analysis would only be valid for the
onset of island formation, it determines the initial placement of
SAQDs; thus determining the initial mean-spacing ($L_0$) and order ($\n$). At later
stages, the SAQDs either order or ripen \cite{Golovin:2003ms,Liu:2003ig,Liu:2003qi,Wang:2004dd,Ross:1998fk}.
The spacing and order established at the small fluctuation stage will
influence the order at a later stage. This has also been verified
through non-linear calculations in Ref. \cite{FriedmanStochastic}. Linear
effects also set the length and the time scale for measuring the perturbations,~\cite{Spencer:1993ve} and determine the arrangement of dots. Linearization
offers a transparent way for analysis and is also a prerequisite for
understanding more advanced non-linear models. The procedure for order
analysis follows Refs.~\cite{FriedmanJNP07,Friedman:fk,FriedmanStochastic}.
Most models in literature are deterministic; however, the stochastic
model is more realistic, as there is no rigorous physical explanation
for the artificial initial random roughness in the deterministic models. 

The rest of this article is organized as follows. We give details
of the stochastic surface diffusion model in section \ref{sec: Deterministic Model}.
In section \ref{sec:order analysis} we discuss the order calculations
using film height correlation functions. We present our
conclusions in section \ref{sec:conclusion}.

\section{Model}

\label{sec: Deterministic Model}

The formation of SAQDs takes place through surface diffusion
that is driven by a diffusion potential $\mu$ and contains thermal fluctuations, $\boldsymbol{\xi}(\x,t)$.~\cite{FriedmanStochastic}. $\mu$ is a non-local
functional of the film height $\h$ and a function of the horizontal position $\x=(x,y)$ (Fig.~\ref{fig:3Dfigurefilmheight}), so that $\mu\rightarrow\mu[\h]\left(\x\right)$.

The normal velocity of the evolving film surface is
\begin{equation}
v_{n}=D\boldsymbol\nabla_{s}^{2}\mu+\boldsymbol\nabla_s \cdot \boldsymbol{\xi}(\x,t),\label{eqn:governing surface diffusion eqn}\end{equation}
where $\boldsymbol\nabla_{s}^2$ is
the surface laplacian, $\boldsymbol\nabla_{s}\cdot$ is the surface divergence, and we omit explicit coordinate and time dependences
for brevity. Here we consider the case of annealing of a film and
therefore we omit a surface flux term in Eq.~\ref{eqn:governing surface diffusion eqn}. 
\label{sec: linearization det model} 

We linearize all quantities about the average film height
$\bar{H}$,
 \begin{equation}
\h(\mathbf{x},t)=\bar{\h}+h(\mathbf{x},t),\end{equation}
where the average film height $\bar{\h}$ can be controlled by controlling
the amount of deposited material and $h(\mathbf{x},t)$ represents
the fluctuations about this average that cannot be experimentally controlled.
In this procedure, the elastic contribution
is non-local, so analysis is aided by working with Fourier components.
Following Refs.~\cite{Spencer:1993ve,FriedmanJNP07} we use the Fourier
transform convention, $f(\x)=\int d^{2}\x\, e^{i\vk\cdot\x}f_{\vk}$
and $f_{\vk}=\left(2\pi\right)^{-2}\int d^{2}\vk\, e^{-i\vk\cdot\x}f(\x)$. Note that subscript $\vk$ is used to indicate functions of wave vector $\vk$ while $(\x)$ is used to indicate dependence on the real-space coordinate.
We proceed in two steps. First, we linearize the diffusion potential
$\mu$. Then, we linearize the dynamic governing equations. Similar to Ref. \cite{FriedmanJNP07}, we keep terms only to linear order
in $h(\mathbf{x},t)$. 

Previously, the \emph{homogenous}  elasticity approximation was used to identify three related wavenumbers and wavelengths,~\cite{Ozkan:1999gf,Obayashi:1998fk}
the characteristic or cutoff wavenumber and wavelength $k_{c}$ and $L_{c}= 2 \pi/k_{c}$, 
the wavenumber and wavelength for maximum energy release, $k_{E}=\left(1/2\right)k_{c}$ and $L_E=2L_c$,~\cite{Ozkan:1999gf,Obayashi:1998fk,Golovin:2003ms}
and the wavenumber and wavelength of the fastest growing mode was identified,
$k_{0}$ and $L_0=2\pi/k_0$.~\cite{Ozkan:1999gf,Obayashi:1998fk} In the absence of
a wetting effect such as for thick films, $k_{0}=\left(3/4\right)k_{c}$ ($L_0=4/3 L_c$),
while for thin films where the wetting effect is strong, $k_{0}$
ranges from $k_{0}=k_{E}$ ($L_0=L_E$) at the critical film height to $k_{0}=(4/3)k_{E}$ ($L_0=(3/4)L_E$).~\cite{Golovin:2003ms,Friedman:fk,FriedmanJNP07}
In the less approximate formulation that is elastically \emph{heterogeneous} and \emph{anisotropic}, these relationships
are not as simple. In the following analysis, we identify $k_{E}$
and $k_{0}$.

\subsection{Energetics}

The diffusion potential, $\mu$, consists of three parts,
$
\mu=\mu_\text{elast.}+\mu_\text{surf}+\mu_\text{wet.}
$.~\cite{Golovin:2003ms,Golovin:2004sa,Tekalign:2004jh,Zhang:2001cr,Friedman:fk}
The elastic energy part destabilizes the two-dimensional growth
made, the surface energy term stabilizes the short wavelength
(high-$k$) modes, and the wetting potential stabilizes all wavelengths.
We proceed by calculating the Fourier transform of the diffusion potential,
$\mu_{\vk}$, to linear order in terms of the Fourier transform of
the film height, $h_{\vk}$.

\subsubsection{Elastic anisotropy and heterogeneity}
\label{sec:elast an het}

The elastic contribution to the diffusion potential is just the elastic
energy density at the film surface, denoted $\omega\left(\x\right)$ times the atomic volume ($\mu_\text{elast.}=\Omega \omega(\x)$).~\cite{Freund:2003ih} We proceed by calculating the Fourier transform of $\omega(\x)$, $\omega_{\vk}$
to linear order in surface height fluctuations, $h_\vk$ while taking into account the effect of both elastic heterogeneity and elastic
anisotropy. 

The full calculation is described in the Appendix, and it results in an elastic energy density of the form\begin{equation}
\omega_{\vk}=-\e_{\theta_{\vk}}(k\bar{H})kh_{\vk},\label{eqn: omegaklin}\end{equation}
 where $\e_{\theta_{\vk}}(k\bar{\h})$ is the elastic energy
density prefactor that depends on both wave vector direction $\theta_{\vk}$
and dimensionless product $k\bar{\h}$. Note that this form is
equivalent to writing $\e\left(\vk\bar{\h}\right)$ where $\vk\bar{\h}$
is a dimensionless vector quantity. This result should be contrasted
with previous calculations. In the \emph{homogeneous isotropic} approximation, the
prefactor is a constant, and in the \emph{homogeneous anisotropic} approximation, the prefactor depends only on the wavevector direction $\theta_{\vk}$.~\cite{Obayashi:1998fk,FriedmanJNP07}

\begin{figure}[!h]
\includegraphics[width=8.5cm,keepaspectratio]{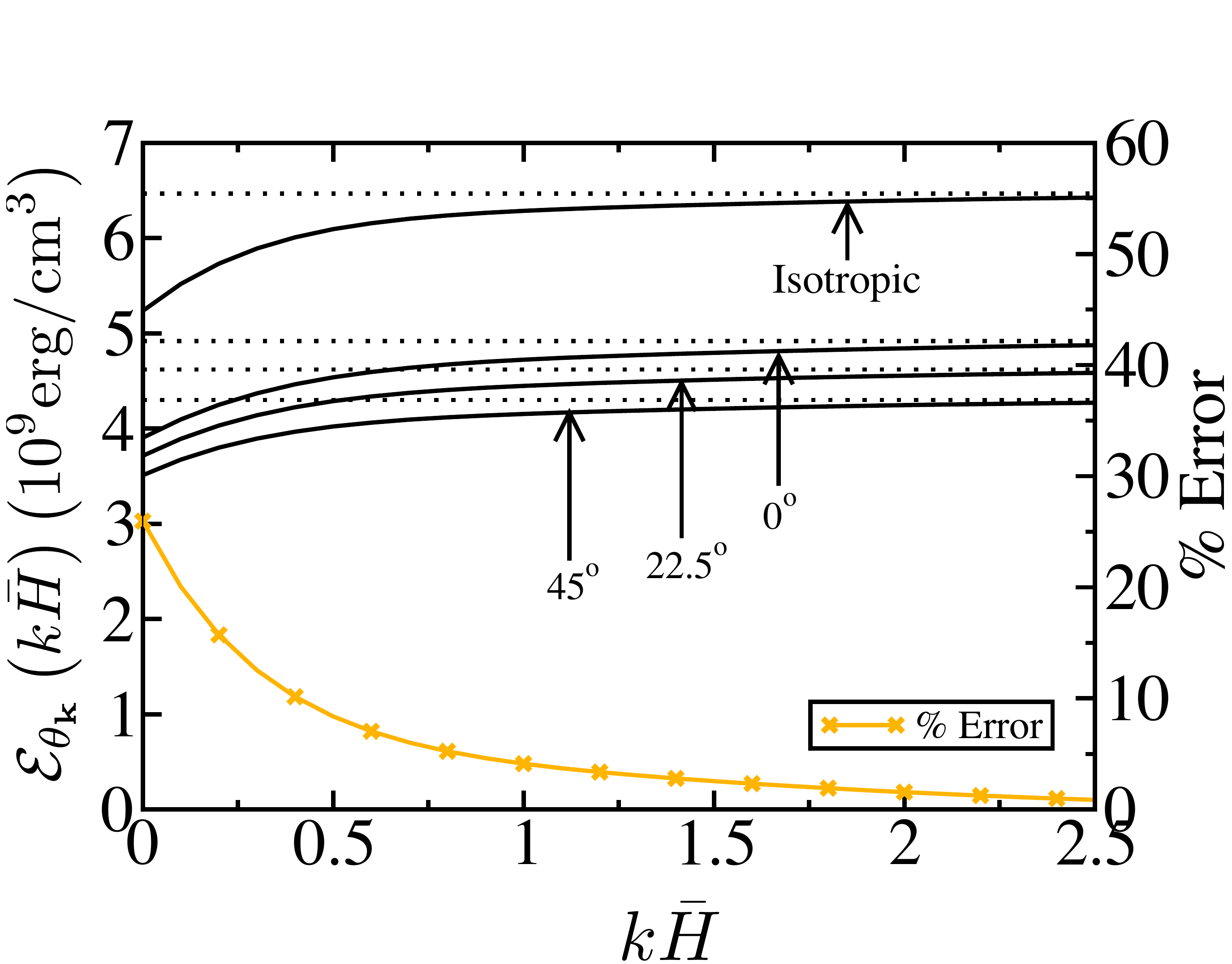}
\caption{Elastic energy density prefactor $\mathcal{E}_{\theta_{\vk}}$ for Ge/Si at $600\,\text{K}$ from \emph{isotropic} approximations~\cite{Spencer:1993ve} and \emph{anisotropic} calculations for $\theta_{\vk} = 0^{\circ}, 22.5^{\circ}\,\text{and}\,45^{\circ}$. Asymptotically large $k\bar{\h}$ limits corresponds to the \emph{anisotropic/isotropic homogeneous} approximations and are shown as dotted lines ($\ldots$). Percent error in the values of $\mathcal{E}_{\theta_{\vk}}$ for $\theta_{\vk} = 0^{\circ}$ is shown  for the \emph{anisotropic homogeneous} approximation.\label{fig:2D prefactor}}
\end{figure}

We perform numerical calculations for [001]-oriented Ge/Si at $600\text{ K}$
to give a concrete example of the energy prefactor $\e_{\theta_{\vk}}\left(k\bar{\h}\right)$.
The elastic stiffness tensor $c_{ijkl}$ is $4-\text{fold}$ symmetric
for rotations about the {[}001] axis; thus $\mathcal{E}_{\theta_{\vk}}\left(k\bar{\h}\right)$
is also $4-\text{fold}$ symmetric. This symmetry manifests itself
in the arrangement of SAQDs into a four-fold symmetric quasiperiodic
lattice \cite{Obayashi:1998fk,Gao:1999ve,Ozkan:1999gf,FriedmanJNP07}.
We use the following physical constants. The superscripts $f$ and
$s$ differentiate between the elastic constants of the film and the
substrate respectively. For Ge at $600\text{ K}$, the elastic constants are $c_{11}^{f}=11.99\times10^{11}\,\text{dyn}/\text{cm}^{2}$,
$c_{12}^{f}=4.01\times10^{11}\,\text{dyn}/\text{cm}^{2}$ and $c_{44}^{f}=6.73\times10^{11}\,\text{dyn}/\text{cm}^{2}$.
\cite{Vorbyev:1996fk} For Si at $600\text{ K}$, $c_{11}^{s}=15.61\times10^{11}\,\text{dyn}/\text{cm}^{2}$,
$c_{12}^{s}=5.63\times10^{11}\,\text{dyn}/\text{cm}^{2}$ and $c_{44}^{s}=7.82\times10^{11}\,\text{dyn}/\text{cm}^{2}$.
\cite{Vorbyev:1996fk} Using $a_{\text{Ge}}=0.5658\,\text{nm}$ and
$a_{\text{Si}}=0.5431\,\text{nm}$, the mismatch strain is $\epsilon_{m}=0.0418$.
Figure~\ref{fig:3D prefactor} shows a plot of the elastic energy
prefactor, $\e_{\theta_{\vk}}\left(k\bar{\h}\right)$ against the
dimensionless variables $k_{x}\bar{\h}$ and $k_{y}\bar{\h}$. Figure~\ref{fig:2D prefactor}
shows $\e_{\theta_{\vk}}\left(k\bar{\h}\right)$ as a function of
$k\bar{\h}$ for three values of $\theta_{\vk}$ along with a comparison
to the discussed homogenous and isotropic approximations. In Figure~\ref{fig:2D prefactor},
we can clearly see that as $k\bar{\h}\rightarrow\infty$, the prefactor
reaches its asymptotic values that also correspond to the homogeneous
approximation.

Typical values for film height are $\bar{\h} < 20\,\text{ML}$, while typical relevant wavelengths are $30-40\text{ nm}$; thus, relevant values for $k\bar{\h}$ are $<1$. In figure~\ref{fig:2D prefactor}, we also plot the error due to the  \emph{homogeneous anisotropic} approximation for  $\theta_{\vk} = 0^{\circ}$ and $0 \leq k\bar{\h} \leq 2.5$. For values of $k\bar{\h} > 2.5$ the error is significantly less (error $< 1\%$).
It should be noted that we focus primarily on the values at $\theta_{\vk}=0^{\circ}$
because undulations are more likely to grow in the $\langle100\rangle$
directions \cite{Obayashi:1998fk,Ozkan:1999gf,FriedmanJNP07}. At lower $k\bar{\h}$ values ($k\bar{\h}<0.4$) the error is higher ($>10\%$) with the upper bound being $26 \%$.  
For example, for a periodic array of islands spaced at $L_{0}=49.06\,\text{nm}\,(k=2\pi/L_{0}=0.128\,\text{nm}^{-1})$
with average film height $\bar{\h}=4.25\,\text{ML}=1.2\,\text{nm}$ so that $k\bar{\h}=0.154$,
the error in the calculation of elastic energy density is about $18 \%$. We find that the \emph{anisotropic thin-film} approximation does a bit better with an error of $-7\%$.
We report these final values along with comparisons to other approximations in Table.~\ref{tab:compare}.
Such errors limit the accuracy of quantitative models, as this error
propagates to calculations of the various characteristic wavenumbers
(Secs.~\ref{subsubsec:energycostfunc} and \ref{subsec:dispersion relation}), mean dot spacing, rate
of growth and critical film height. %

\begin{table*}
\caption{
Comparison of presented model with various approximations.  Our model uses \emph{heterogeneous anisotropic} elasticity (Het. Anis.).  Other models use \emph{homogeneous anisotropic} elasticity (Hom. Anis.)~\cite{Obayashi:1998fk,Ozkan:1999gf}, \emph{anisotropic thin-film} elasticity (Anis. Thin), \emph{heterogeneous isotropic} elasticity (Het. Iso.)~\cite{Spencer:1993vt}, \emph{homogeneous isotropic} elasticity (Hom. Iso.) and \emph{isotropic thin-film} elasticity (Iso. Thin)~\cite{Tekalign:2007}  All calculations use average film height $\bar{\h}=4.25\text{ ML}$ and $\e_{0\degree}(0.154)$ uses $k = 2\pi/ (49.06\text {nm}) = 0.128\text{ nm}^{-1}$. Values in parentheses indicate \% error due to each approximation.
}
\begin{center}
\begin{tabular}{|c|c|c|c|c|c|c|}
\hline
&Het. Anis.& Hom. Anis. & Anis. Thin & Het. Iso. & Hom. Iso. & Iso. Thin\\
& &(\% error)&(\% error)&(\% error)&(\% error)&(\% error)\\ 
\hline
$\e_{0\degree}(0.154)$ &$4.18$ &$4.92$ &$3.90$ &$5.64$ &$6.47$ &$5.24$\\
($10^9$ erg/cm$^3$) & &$(+18\%)$ &$(-7\%)$ &$(+35\%)$ &$(+56\%)$ &$(+25\%)$\\
\hline
$L_E$ (nm) &$55.4$ &$49.2$ &$62.1$ &$39.9$ &$37.4$ &$46.2$\\
& & $(-11\%)$ &$(+12\%)$ &$(-28\%)$ &$(-32.5\%)$ &$(-17\%)$\\
\hline
$L_0$ (nm) &$49.1$ &$43.7$ &$55.1$ &$35.4$ &$33.2$ &$41.0$\\
& &$(-11\%)$&$(+12\%)$ &$(-28\%)$ &$(-32\%)$ &$(-17\%)$\\
\hline
$n_\text{cor.}$ &$2.71$ &$2.07$ & $2.62$ &n/a &n/a &n/a\\
& &$(+24\%)$ & $(-3\%)$ & & &\\
\hline

\end{tabular}
\end{center}
\label{tab:compare}
\end{table*}
\subsubsection{Surface and Wetting energies}

The other contributions to the SAQD formation energetics are the surface
and wetting energies. Since our focus is on $4\text{--}$fold symmetric
systems, the only anisotropic term is due to the elastic energy.~\cite{FriedmanJNP07} As in Ref. \cite{FriedmanJNP07},
\begin{equation}
\mu_{\text{surf.,}\vk}=\Omega(\gamma k^{2})h_{\vk},\label{eqn:muklin surface energy}\end{equation}
 where $\gamma$ can be interpreted as the effective surface energy.~\cite{Gao:1999ve,FriedmanJNP07}
The linearized wetting potential is \begin{equation}
\mu_{\text{wet},\vk}=\Omega\left(W^{\prime\prime}\right)h_{\vk},\label{eqn:muklin wetting energy}\end{equation}
 where $W^{\prime\prime}$ is the second derivative of the wetting
potential with respect to the film height evaluated at the average
film height $\h=\bar{\h}$. For the example here, we follow Ref.~\cite{Zhang:2001cr}
and take the wetting potential to be $W=B/\h$, where $B$ is a material
constant. 

\subsubsection{Energy cost function}
\label{subsubsec:energycostfunc}

\begin{figure}[!h] 
\centering\includegraphics[width=8.5cm,keepaspectratio]{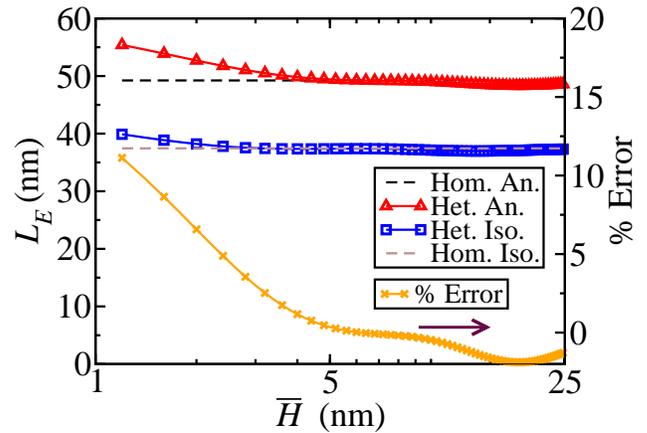}
\caption{Characteristic wavelength $L_{E}$ as function of $\bar{H}$ for Ge/Si at $600\,\text{K}$ from \emph{isotropic} and \emph{anisotropic} calculations. Corresponding \emph{homogeneous} approximations are shown as dashed lines. The percent error in  is shown for the \emph{anisotropic  homogeneous} approximation.\label{fig:L_E}}
\end{figure}

Combining Eqs.~\ref{eqn: omegaklin}, \ref{eqn:muklin surface energy}
and \ref{eqn:muklin wetting energy}, we can write the linearized
diffusion potential in Fourier space as \begin{equation}
\mu_{\vk}=f(k,\theta_{\vk},\bar{H})h_{\vk},\end{equation}
 where $f(k,\theta_{\vk},\bar{H})=\Omega\left[-k\mathcal{E}_{\theta_{\vk}}(k\bar{H})+\gamma k^{2}+W^{\prime\prime}\right]$
is the energy cost per unit height for a periodic perturbation. The
minima in the energy cost function lie along the $\left\langle 100\right\rangle $ directions
($\theta_{\vk}=0^{\circ},\,90^{\circ},\,180^{\circ},\,270^{\circ}$)
and occur at wavenumber $k_{E}$ so that 
the most energy is released when dots form at a period of $L_{E}=2\pi/k_{E}$.
$L_{E}$ is a function of $\bar{\h}$, a dependence that is due purely
to the more precise elasticity calculation we present, and not, for
example, a result of the wetting potential. For a concrete example,
we use the estimated surface energy density $\gamma=1927\,\text{erg}/\text{cm}^{2}$
and the atomic volume $\Omega=2.27\times10^{-23}\,\text{cm}^{3}$.
Figure~\ref{fig:L_E} shows $L_{E}$ as a function of $\bar{\h}$ along with
its values for the \emph{anisotropic homogenous} and \emph{isotropic} approximations. As $\bar{\h}$ becomes
large, $L_{E}$ approaches the \emph{anisotropic homogeneous} approximation value that
is independent of $\bar{\h}$. For an average film height, $\bar{\h} = 4.25\,\text{ML}=1.2\,\text{nm}$, the error
in $L_{E}$ from the homogeneous approximation is about $11 \%$.  We report results for all approximations in Table.~\ref{tab:compare}.

The energy cost function is also useful for determining the critical
film thickness for SAQD formation, or for modeling purposes it can
be used to estimate the wetting potential that leads to an observed
critical height. The critical height for SAQD formation in the Ge/Si film-substrate system is generally observed to be $4\text{--}6$ ML. \cite{Hegazy:2006} Here, we choose a critical film height of $4\,\text{ML}$ and follow the procedure from Ref.~\cite{Zhang:2001cr}. We assume a wetting potential of the form $W(\h)=B/\h$ and then find the coefficient $B$ that gives a critical film height, $H_{c}=4\,\text{ML}$, by setting the minimum value for the energy cost function
to zero, $f_{\text{min}}=f(k_{E},0^{\circ},\h_{c})=0$. Solving
for $B$ is a simple procedure as $f$ is linear in $B$. We find that even the value of $B$ is sensitive to the \emph{anisotropic homogeneous} and other approximations. At 4~ML, $B=1.61\times10^{-6}\,\text{erg}/\text{cm}$ for the full theory, and $B=2.28\times10^{-6}\,\text{erg}/\text{cm}$ for
the \emph{anisotropic homogeneous} approximation, about $42\%$ difference.%

\subsection{Dispersion Relation}

\label{subsec:dispersion relation}

The linearized evolution equation in Fourier space is given by
\cite{FriedmanStochastic} \begin{equation}
\partial_{t}h_{\vk}=\sigma_{\vk}h_{\vk}+\sqrt{2\Omega Dk_{b}T}\left[i\vk\cdot\boldsymbol\eta_{\vk}(t)\right],\label{eq:se}\end{equation}
where the second term is the Fourier transform of $\boldsymbol\xi(\x,t)$ to linear order, 
$\langle\boldsymbol\eta_{\vk}(t) \boldsymbol\eta_{\vk'}^*(t') \rangle
=(2\pi)^{-2}\tilde{I}\delta^2(\vk-\vk')\delta(t-t')
$
,~\cite{FriedmanStochastic} and \begin{equation}
\sigma_{\vk}=-Dk^{2}f(k,\theta_{\vk},\bar{H})\end{equation}
is the generalized dispersion relation that gives
the rate of growth (positive values) or decay (negative values) of
each height Fourier component $h_{\vk}$.

\begin{figure}[htb] 
\centering\includegraphics[width=8.5cm,keepaspectratio]{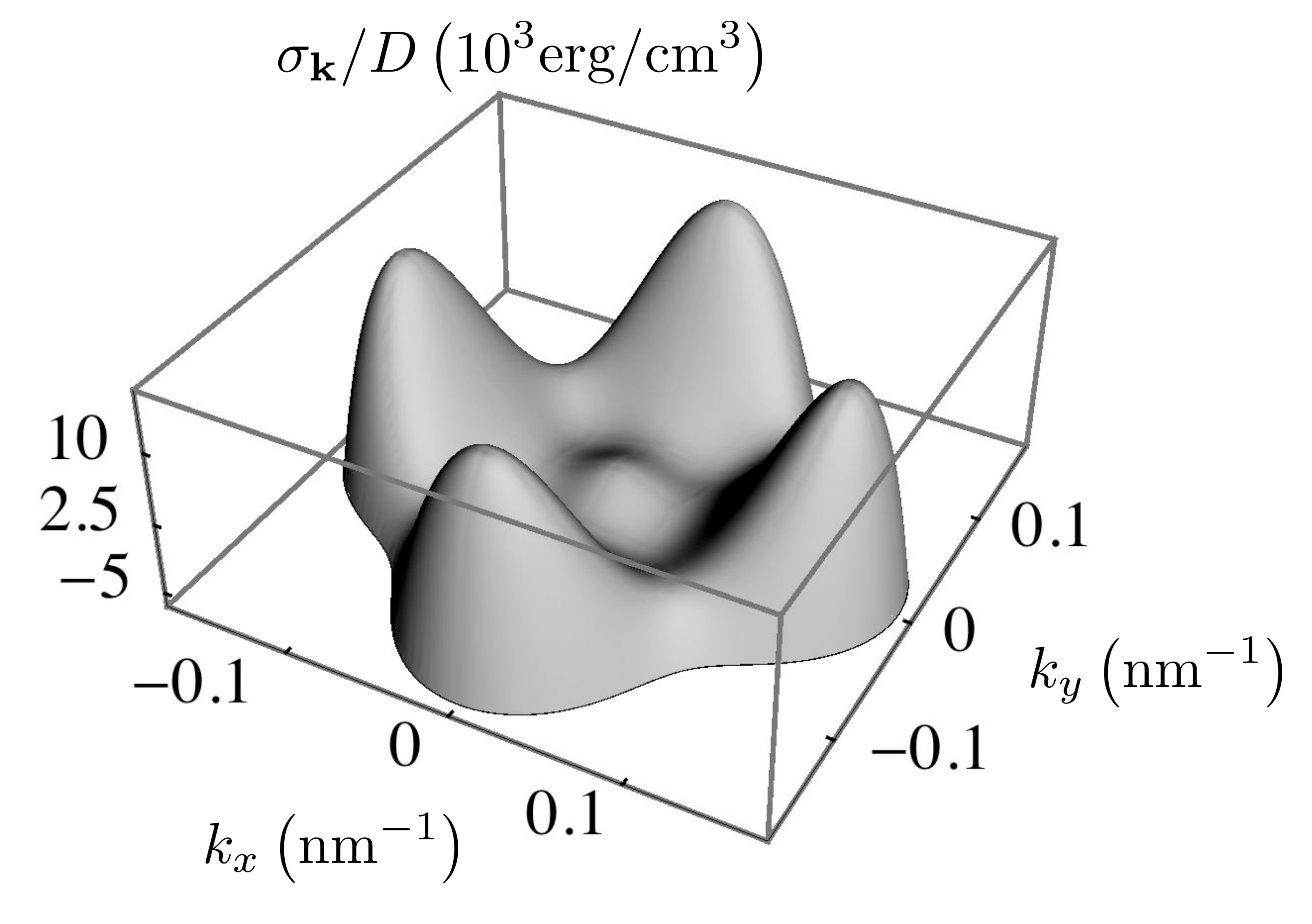}
\caption{$\sigma_{\vk}/D$ vs $\vk$ for $\bar{\h}=4.25\,\text{ML}$ for Ge/Si at $600\,\text{K}$.\label{fig:sigma_k}}
\end{figure}

Figure~\ref{fig:sigma_k} shows the dispersion relation $\sigma_\vk$
for $\bar{\h}=4.25\,\text{ML}$. For the case shown in figure~\ref{fig:sigma_k},
$\sigma_\vk$ has $4$ peaks along the $\langle100\rangle$
directions at $\vk_0=(0, \pm0.128)\,\text{nm}^{-1}$ and $(\pm0.128,0)\,\text{nm}^{-1}$.
The four peaks indicate
that the instability is maximum in the $\langle100\rangle$ directions
thus making them the likely directions for the alignment of SAQDs.
This alignment is consistent with previous studies.~\cite{Obayashi:1998fk,Ozkan:1999gf,Zhang:2003tg,FriedmanJNP07}

Similar to Ref. \cite{FriedmanJNP07}, we expand $\sigma_{\vk}$
about its peak values to get \begin{equation}
\sigma_{n}=\sigma_{0}-\frac{1}{2}\sigma_{\parallel}(k-k_{\parallel})^{2}-\frac{1}{2}\sigma_{\perp}k_{\perp}^{2},\label{eq:thesigma}\end{equation}
 where \begin{equation}
\sigma_{\parallel}=-\left.\frac{\partial^{2}\sigma_{\vk}}{\partial k^{2}}\right|_{\left[k_{0},(\theta_{0})_{n}\right]},\quad\sigma_{\perp}=-\frac{1}{k_{0}^{2}}\left.\frac{\partial^{2}\sigma_{\vk}}{\partial\theta^{2}}\right|_{\left[k_{0},(\theta_{0})_{n}\right]},
\label{eq:sigmas}
\end{equation}
 $n$ corresponds to the number of peaks, $\theta_{0}$ is the orientation
of $\vk_0$, and $k_{\parallel}$ and $k_{\perp}$ are components
parallel and perpendicular to $\vk_0$. 
We discuss the dependence of $k_{0}$ and the mean dot spacing on
film thickness next along with the discussion of SAQD array order.

\section{Order Analysis}

\label{sec:order analysis} 

The spatial order of SAQDs is best characterized by the mean geometric
spacings, $L_{0}$, and alignments and by the degree of fluctuation about these
means. The average alignment of SAQDs is $\left\langle 100\right\rangle $, and we characterize the range of order by, $\n$, the number of
dots in a row whose positions are likely to be well correlated, meaning
that they are likely to be both regularly spaced and well-aligned. In the following discussion
we present calculations for different average film heights, and for each film height we calculate average dot spacing and the number of correlated dots when film height
fluctuations reach atomic scale size. For the second part, for finding
$\n$, we use the film height correlation function and associated correlation
lengths which were derived previously.~\cite{Friedman:fk,FriedmanJNP07}

\subsection{Average Dot Spacing}
\label{subsection:meandotspacing}

\begin{figure}[!h] 
\centering
\includegraphics[width=8.5cm,keepaspectratio]{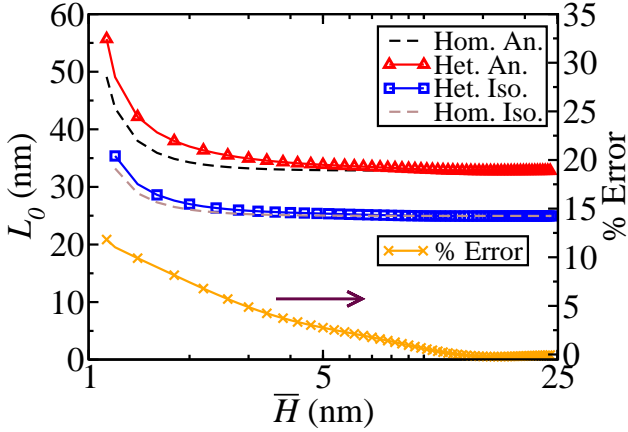}
\caption{Average dot spacing $L_{0}$ as function of $\bar{H}$ for Ge/Si at $600\,\text{K}$ from \emph{isotropic} and \emph{anisotropic} calculations. Corresponding \emph{homogeneous} approximations are shown as dashed lines. The percent error in  is shown for the \emph{anisotropic  homogeneous} approximation.\label{k_0Sigmak}}
\end{figure}

As done previously,~\cite{Spencer:1993vt,Golovin:2003ms,Friedman:fk,FriedmanJNP07,FriedmanStochastic} we estimate the average initial spacing between dots to be $L_{0}=2\pi/k_{0}$.
Figure~\ref{k_0Sigmak} shows a plot of $L_{0}$ against $\bar{\h}$
and compares it with the results for the \emph{anisotropic homogeneous} and \emph{isotropic}
approximations. The error associated with the homogeneous approximation
is also shown. We report values for $\bar{\h}=1.2\text{ nm}$ in Table~\ref{tab:compare}.  For the example studied here (Ge/Si at $600\text{ K}$),
the value of average spacing for \emph{anisotropic heterogeneous} elasticity calculation
varies between $32.8\,\text{nm}$ to $55.7\,\text{nm}$. Note that
as $\bar{\h}\rightarrow\infty$, $L_{0}$ reaches its asymptotic value,
which is the same as the value from the \emph{anisotropic homogeneous} approximation.
Thus, the \emph{anisotropic homogeneous} approximation is valid for thick films. Typically
experiments correspond to values of $\bar{\h}$ that are less than
$20\,\text{ML} (4.25\,\text{nm})$. 

\subsection{Order Analysis Using Correlation Functions}

The autocorrelation function and its Fourier transform, also known
as the power spectrum, are very useful for characterizing dot order.~\cite{Stangl:2000superlattice,Springholz:2001nx,FriedmanJNP07} The
autocorrelation function is defined as \[
C^{A}(\dx)=\int d^{2}\x\, h(\x+\dx)h(\x).\] For
an imperfectly periodic array of SAQDs the autocorrelation function
decays away from the origin. The distance over which the autocorrelation
function decays is known as the correlation length $L_{\text{cor}}$. The value $2L_{\text{cor}}$ represents the distance over which the
SAQDs appear to be periodic meaning regularly spaced and well aligned.

The power spectrum is \[
C_{\vk}^{A}=\left|h_{\vk}\right|^{2}.\] 
The power spectrum for a nearly periodic array of SAQDs will have
peaks with finite width, $\Delta_{k}$. 
The spectrum peak width $\Delta_{k}$  is related to the correlation length $L_{\text{cor}}$
by $L_{\text{cor}}=1/\Delta_{k}$.

Each simulation or experiment corresponds to one particular realization
with its own autocorrelation function; however, for sufficiently large
simulation sizes, the fluctuations in $C^{A}\left(\dx\right)$ are
small, and the ensemble average of the autocorrelation functions can
be predicted and provides a good estimate of individual autocorrelation
functions and spectrum functions.~\cite{FriedmanJNP07} The ensemble
average of the autocorrelation function is the correlation function
$C(\dx)=\left\langle C^{A}(\dx)\right\rangle =\left\langle h(\dx)h(\mathbf{0})\right\rangle $.
Similarly, the ensemble average spectrum function is $C_{\vk}=\left\langle C_{\vk}^{A}\right\rangle $,
where $C_{\vk}$ is also the Fourier transform of $C(\dx)$, and $C_{\vk}$
is the coefficient in the covariance of the Fourier components $h_{\vk}$;
$\left\langle h_{\vk}h_{\vk'}^{*}\right\rangle =C_{\vk}\delta^{2}(\vk-\vk')$, where $\delta^{2}(\vk-\vk')$ is the two-dimensional Dirac Delta function.

The spectrum function can be solved using Eqs.~\ref{eq:se}, \ref{eq:thesigma} and~\ref{eq:sigmas},~\cite{FriedmanStochastic}
\begin{equation}
C_{\vk}\approx\frac{D\Omega k_{b}T}{(2\pi)^{2}\sigma_{0}}k^{2}e^{2\sigma_{0}t}\sum_{n=1}^{4}e^{-\frac{1}{2}L_{\parallel}^{2}(k_{\parallel}k_{0})^{2}-\frac{1}{2}L_{\perp}^{2}k_{\perp}^{2}},
\end{equation}
where $L_{\parallel}=\sqrt{2\sigma_{\parallel}t}$ and $L_{\perp}=\sqrt{2\sigma_{\perp}t}$
are the correlation lengths. $L_{\parallel}$ gives about half the
length over which the dot spacing is regular, while $L_{\perp}$ gives
about half the length over which a row of dots is straight. Of the
two correlation lengths, $L_{\perp}$ tends to be smaller and thus
more limiting.
Taking the inverse Fourier transform, the correlation function
is \begin{multline}
C(\mathbf{\Delta x})\approx\frac{D\Omega k_{b}Tk^{2}}{\pi\sigma_{0}L_{\parallel}L_{\perp}}e^{2\sigma_{0}t}\left[e^{-\frac{1}{2}(\Delta x^{2}/L_{\parallel}^{2}+\Delta y^{2}/L_{\perp}^{2})}\cos(k\Delta x)\right.\\
\left. +e^{-\frac{1}{2}(\Delta x^{2}/L_{\perp}^{2}+\Delta y^{2}/L_{\parallel}^{2})}\cos(k\Delta y)\right].\end{multline}

\begin{figure}[!h] 
\centering\includegraphics[width=8.5cm,keepaspectratio]{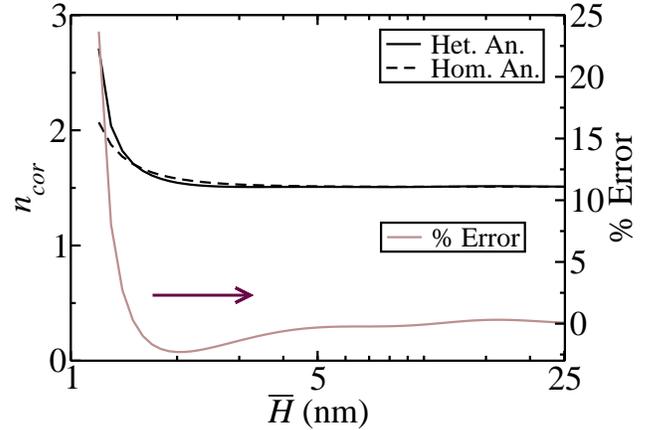}
\caption{Number of correlated dots $\n$ for small height fluctuations Ge/Si at $600\,\text{K}$ for \emph{anisotropic heterogeneous} and \emph{anisotropic homogeneous} calculations. Also shown, error in $\n$  from \emph{anisotropic homogeneous} approximation.\label{fig:stochastic no of dots}}
\end{figure}

Figure~\ref{fig:stochastic no of dots} shows the number of correlated
dots calculated as $\n=2L_{\perp}/L_{0}$ for the small fluctuation
stage ($C(\mathbf{\Delta x}=\mathbf{0})=1\,\text{ML}^{2}$). Both error and number of correlated dots decline sharply for a small increment in film height above the critical film height. We find the error drops from $24 \%$ at $\bar{\h} = 4.25\,\text{ML}$ to $3 \%$ at $\bar{\h} = 4.95\,\text{ML}$. Note that with further increase in $\bar{\h}$, the error fluctuates between $\pm 2 \%$ before reaching $0 \%$ for higher $\bar{\h}$ values. 

\section{Conclusion}

\label{sec:conclusion}

Most theoretical and many numerical models of SAQD growth approximate
film-substrate systems as elastically homogeneous. We have examined
the effect of elastic heterogeneity on SAQD mean spacing and the order
estimates developed in Refs.~\cite{FriedmanJNP07} and~\cite{FriedmanStochastic}.
We have performed a linear analysis incorporating both elastic heterogeneity
and elastic anisotropy. We quantify the effect of heterogeneity
as percent error in the calculated values of elastic energy density, average spacing between the SAQDs and number of correlated dots based on homogeneous approximation.
We show that the homogeneous approximation of the film-substrate system
can lead to significant errors in the calculations for formation and
ordering of SAQDs. For the case of Ge/Si system at $600\,\text{K}$
the upper bound for error in the calculated value of elastic energy density
is found to be as large as $26\%$. The error declines as $\bar{\h}$ increases.  For a typical average film height, $\bar{\h}=4.25\,\text{ML}$, we calculate an error of about $11\%$
in the estimation of average spacing between the SAQDs. 
Using a stochastic model, and the film height correlation functions, we
find that the error in the estimated number of correlated dots declines quickly as $\bar{\h}$ increases. The error in the estimated number of correlated dots drops from about $24\%$ at $\bar{\h} = 4.25\,\text{ML}$ to about $3\%$ at $\bar{\h} = 4.95\,\text{ML}$. 
For thinner films, the thin-film approximations can reduce this error, but error still remains as the thin-film approximation actually overestimates the effect of elastic heterogeneity.
In general we find that the most error is due to using an isotropic approximation.  
For the isotropic heterogeneous approximation~\cite{Spencer:1993vt} the error in mean dot spacing remains more or less constant at $34\%$ for values of $\bar{\h} < 20\,\text{ML}$. We did not report order predictions from isotropic models, as they are inappropriate for order and alignment estimates.~\cite{Friedman:fk}

The interplay between elastic strain, surface energy and surface diffusion
can be quite complicated. Errors introduced by the elasticity portion
of models can confound our ability to asses how well we model surface
and wetting energy. Given the challenge of developing accurate models
of surface and wetting energies, it is essential that the elasticity
part of the calculation be correct. Here we have used a linear model
that includes elastic anisotropy and heterogeneity. This is an important
step in the development of a complete non-linear stochastic model
required for a more comprehensive quantitative analysis, for example
incorporating surface energy and diffusive anisotropy. We have demonstrated
that one must incorporate both elastic anisotropy
and elastic heterogeneity to avoid the introduction of significant errors in the calculation of the elastic energy density and mean dot spacing. This
effect will have important consequences for single layer arrays grown
on a flat [001] surface, but should also suggest the magnitude of
errors introduced in modeling other SAQD morphologies such as multilayers
and growth on patterned substrate.

\section*{Acknowledgment}
C. Kumar gratefully acknowledges financial support The Pennsylvania State University Graduate Fellowship program.

\appendix

\section*{Appendix: Elastic Energy Coefficient\label{appendix:elasticity calculations}}

The increase in elastic energy due to the addition of a small material
volume at the surface is just the elastic energy density at the surface.~\cite{Freund:2003ih}
We calculate the elastic energy density using perturbation theory following
Ref.~\cite{FriedmanJNP07}, but here we take elastic heterogeneity
into account. This calculation was performed previously but only with
the approximation of isotropic elasticity.~\cite{Spencer:1993ve,Tekalign:2004jh}
Here we incorporate both elastic heterogeneity and elastic anisotropy
to calculate the elastic energy density at the film surface.

We consider a flat film on a substrate. The lattice mismatch between
the film and the substrate introduces a misfit strain $\epsilon_{m}$
in the film and leads to a uniform stress distribution in the film
given by~\cite{Freund:2003ih,FriedmanJNP07} \begin{equation}
\tilde{\sigma}_{m}=\begin{bmatrix}\sigma_{m} & 0 & 0\\
0 & \sigma_{m} & 0\\
0 & 0 & 0\end{bmatrix},\end{equation}
 where $\sigma_{m}=M\epsilon_{m}$, and $M$ is the biaxial modulus,
\begin{equation}
M=\left[c_{11}^{f}+c_{12}^{f}-2(c_{12}^{f})^{2}/c_{11}^{f}\right],\end{equation}
 where $c_{11}^{f}$ and $c_{12}^{f}$ are elastic constants for the
film. We perturb the film surface so the film height fluctuates as
\begin{equation}
h(\mathbf{x})=h_{0}e^{ikx},\end{equation}
where the cartesian coordinate system is set up so that $z=0$ or
the $x\text{--}y$ plane lies at the interface of the film-substrate
system, and the $x-$direction is aligned along $\vk$. Then, we calculate
the elastic energy to first order in the perturbation amplitude, $h_{0}$.
This calculation requires four steps. First we must find the surface
normal vector $\mathbf{n}$ to first order in $h_{0}$. Then, we must
find the admissible equilibrium eigenmodes for the elastic displacement
that have the same periodicity as the height perturbation. Then we
must find the eigenmode coefficients from the surface boundary conditions
and internal matching conditions (compatibility and equilibrium).
Finally, we find the elastic energy density at the free surface to
first order in $h_{0}$. The first step is simple, to the first order
in $h_{0}$ the normal to the surface of the film is given by \begin{equation}
\mathbf{n}(\mathbf{x})=-ikh_{0}e^{ikx}\mathbf{e}_{x}+\mathbf{e}_{z}.\end{equation}
The remaining three steps follow.

We find the elastic deformation eigenmodes that have the same periodicity
as the perturbation, and that satisfy internal equilibrium; working
with the displacement field $\mathbf{u}$ automatically satisfies
compatibility away from internal interfaces. We first construct the
rank $4$ elastic stiffness tensor for both film and the substrate
for an arbitrary passive rotation $\theta_{\vk}$ in the $x\text{--}y$
plane so that $\vk$ can lie along any direction in the plane. \begin{equation}
c_{qrst}^{f}(\theta_{\vk})=\sum_{i,j,k,l=1}^{3}R(\theta_{\vk})_{qi}R(\theta_{\vk})_{rj}R(\theta_{\vk})_{sk}R(\theta_{\vk})_{tl}c_{ijkl}^{f}\end{equation}
 and \begin{equation}
c_{qrst}^{s}(\theta_{\vk})=\sum_{i,j,k,l=1}^{3}R(\theta_{\vk})_{qi}R(\theta_{\vk})_{rj}R(\theta_{\vk})_{sk}R(\theta_{\vk})_{tl}c_{ijkl}^{s},\end{equation}
where we are using the superscript $f$ and $s$ for the film and
the substrate respectively and $R(\theta_{\vk})$ is the passive
rotation matrix by an angle $\theta_{\vk}$ about the $z-$axis. To
match boundary conditions in a later step, displacements must have
the form \begin{align}
u_{l}^{f}(x,y,z) & =U_{l}e^{k(ix+\kappa z)}\label{eq:uf}\\
u_{l}^{s}(x,y,z) & =V_{l}e^{k(ix+\zeta z)}.\end{align}
where $\kappa$ and $\zeta$ are unknown eigenvalues. The stress tensors
in the film and the substrate are \begin{align}
\sigma_{qr}^{f} & =\sum_{s,t=1}^{3}c_{qrst}^{f}\frac{\partial u_{s}^{f}}{\partial x_{t}}+(\tilde{\sigma}_{m})_{qr}\label{eqn: stress state in the film}\\
\sigma_{qr}^{s} & =\sum_{s,t=1}^{3}c_{qrst}^{s}\frac{\partial u_{s}^{s}}{\partial x_{t}}.\end{align}
 For the film, the elastic equilibrium equations are \begin{align}
 & \sum_{q,s,t=1}^{3}\frac{\partial}{\partial x_{q}}c_{qrst}^{f}(\theta_{\vk})\frac{\partial}{\partial x_{s}}u_{t}^{f}=0;n=1\ldots3\\
 & \left(\sum_{t=1}^{3}C_{rt}^{f}(\theta_{\vk},\kappa)U_{t}\right)k^{2}e^{k(ix+\kappa z)}=0,\label{eqn:eigenvalue eqn}\end{align}
 where \begin{equation}
C_{rt}^{f}(\theta_{\vk},\kappa)=\sum_{q,s=1}^{3}c_{qrst}^{f}(\theta_{\vk})(i\delta_{q1}+\delta_{q3}\kappa)(i\delta_{s1}+\delta_{s3}\kappa).\label{eq:crt}\end{equation}
 To obtain non-trivial solutions, we set the determinant of $C_{rt}^{f}(\theta_{\vk},\kappa)$
to zero. We thus obtain six eigenvalues of $\kappa$ denoted by $\kappa^{p}$
with $p=1\ldots6$. Each value of $\kappa=\kappa^{p}$ is substituted
back into $C_{rt}(\theta_{\vk},\kappa)$, and Eq.~\ref{eqn:eigenvalue eqn}
is solved to find the corresponding eigenvectors $U_{l}^{p}$. The
displacement components for the film in terms of the unknown coefficients
$A_{p}$ are thus \begin{equation}
u_{l}^{f}=ih_{0}\epsilon_{m}\sum_{p=1}^{6}A_{p}U_{l}^{p}e^{k(ix+\kappa^{p}z)},\label{eq:uf1}\end{equation}
 where we assume that the perturbing elastic field displacement components
are proportional to $\epsilon_{m}$ and $h_{0}$, and we put the prefactor
$i=\sqrt{-1}$ in for convenience. We use the same procedure to find
the eigenvalues and eigenvectors for the substrate displacements $u_{l}^{s}$,
where $C_{rt}^{s}(\theta_{\vk},\zeta)$ has the same form as Eq.~\ref{eq:crt},
but using the substrate elastic constants, $c_{qrst}^{s}$. Six eigenvalues,
$\zeta^{p}$ are obtained; however, we assume that the substrate is
a semi-infinite solid so the displacement field $u_{l}^{s}=0$ at
$z=-\infty$. Thus, we only retain the three eigenvalues with $\text{Re}\left[\zeta^{p}\right]>0$
that satisfy this condition and discard the other three. We find the
displacement components of the substrate, \begin{equation}
u_{l}^{s}=ih_{0}\epsilon_{m}\sum_{p=1}^{3}B_{p}V_{l}^{p}e^{k(ix+\zeta^{p}z)},\end{equation}
 where $V_{l}^{p}$ are the eigenvectors, and $B_{p}$ are the unknown
coefficients. 

We now find the nine unknown coefficients ($A_{p}$ and $B_{p}$)
using the traction-free boundary condition at the surface, and the
internal matching conditions at the film-substrate interface, namely
equilibrium and compatibility. The traction on the surface of the
film is \begin{equation}
T_{r}=\sum_{q=1}^{3}\sigma_{qr}^{f}n_{q},\label{eqn:traction}\end{equation}
 where $z=\bar{H}+h(\mathbf{x})$. Substituting Eq.~\ref{eqn: stress state in the film}
into Eq.~\eqref{eqn:traction}, we get \begin{align}
T_{r} & =\sum_{q,s,t=1}^{3}\left[c_{qrst}^{f}(\theta_{\vk})\frac{\partial u_{s}}{\partial x_{t}}\right]n_{q}+(\tilde{\sigma}_{m})_{rq}n_{q}\label{eqn: traction film surface}\\
 & =i\epsilon_{m}h_{0}\sum_{q,s,t=1}^{3}\sum_{p=1}^{6}c_{qrst}^{f}kA_{p}U_{s}^{p}(i\delta_{t1}\ldots \nonumber\\
 &\ldots+\kappa^{p}\delta_{t3})e^{k(ix+\kappa^{p}z)}n_{q}+(\tilde{\sigma}_{m})_{rq}n_{q}.\nonumber \end{align}
Again, we substitute $z=\bar{H}+h(\mathbf{x})$, and we keep only
terms up to first order in $h_{0}$ to get 
\begin{equation}
\begin{split}
T_{r}&=\left[\sum_{s=1}^{3}\sum_{p=1}^{6}\left(ic_{3rs1}^{f}(\theta_{\vk})+\kappa^{p}c_{3rs3}^{f}(\theta_{\vk})\right)A_{p}\right.\ldots\\
&\ldots\left.\times U_{s}^{p}e^{k\kappa^{p}\bar{H}}-M\delta_{r1}\right]ik\epsilon_{m}h_{0}e^{ikx}.
\end{split}
\end{equation}
 Since the traction on the film surface must be zero, we have \begin{equation}
\sum_{s=1}^{3}\sum_{p=1}^{6}\left(ic_{3rs1}^{f}(\theta_{\vk})+\kappa^{p}c_{3rs3}^{f}(\theta_{\vk})\right)A_{p}U_{s}^{p}e^{\kappa^{p}k\bar{H}}=M\delta_{r1}\label{eqn:first 3 elasticity eqns}\end{equation}
 giving three equations for $r=1\dots3$. 

The force balance at the internal film-substrate interface requires
\begin{equation}
\left.\sigma_{3r}^{f}=\sigma_{3r}^{s}\right|_{z=0}.\label{eqn: stress balance equation}\end{equation}
 In terms of the unknown coefficients $A_{p}$ and $B_{p}$, we can
write Eq.~\eqref{eqn: stress balance equation} as 
\begin{equation}\label{eqn: 3 more elasticity eqns}
\begin{split}
\sum_{s=1}^{3}\sum_{p=1}^{6}\left(ic_{3rs1}^{f}+\kappa^{p}c_{3rs3}^{f}\right)A_{p}U_{s}^{p}&=\sum_{s=1}^{3}\sum_{p=1}^{3}\left(ic_{3rs1}^{s}\right.\ldots\\
\ldots&\left.+\zeta^{p}c_{3rs3}^{s}\right)B_{p}V_{s}^{p}
\end{split}
\end{equation}
 for $r=1\dots3$. For the compatibility between the film and the
substrate at the interface, the displacements of the film and the
substrate must be equal, so that \begin{equation}
\left.u_{q}^{f}=u_{q}^{s}\right|_{z=0}.\end{equation}
 In terms of the unknowns, the compatibility equation can be written
as \begin{equation}
\sum_{p=1}^{6}A_{p}U_{i}^{p}=\sum_{q=1}^{3}B_{q}V_{i}^{q}\label{eqn:the last 3 elasticity equation}\end{equation}
giving three equations $i=1\dots3$. We then calculate the nine coefficients,
$A_{p}(\theta_{\vk},k\bar{H})$ with $p=1\dots6$ and $B_{q}(\theta_{\vk},k\bar{H})$
with $q=1\dots3$ using using Eqs.~\eqref{eqn:first 3 elasticity eqns},
\eqref{eqn: 3 more elasticity eqns} and \eqref{eqn:the last 3 elasticity equation}. 

Following Ref.~\cite{FriedmanJNP07}, the we find the elastic energy
at the film surface to first order in $h_{0}$ to be
\begin{equation}
U=U_{0}+M\epsilon_{m}\left.\left(\partial_{x_{1}}u_{1}^{f}+\partial_{x_{2}}u_{2}^{f}\right)\right|_{z=\bar{\h}},
\end{equation}
where $U_{0}$ is the energy of the unperturbed flat film, a constant.
Using Eq.~\ref{eq:uf1} \begin{equation}
U=U_{0}-\e_{\theta_{\vk}}(k\bar{H})kh_{0}e^{ikx}\text{, where}\end{equation}
\begin{equation}
\e_{\theta_{\vk}}(k\bar{H})=M\epsilon_{m}^{2}\sum_{p=1}^{6}A_{p}(\theta_{\vk},k\bar{H})U_{1}^{p}(\theta_{\vk}),\end{equation}
where we note that $A_{p}$ will depend on $\theta_{\vk}$, $k$ and
$\bar{\h}$, and $U_{l}^{p}$ will depend on $\theta_{\vk}$. By the
principle of superposition, we can use the elastic energy coefficient
$\e_{\theta_{\vk}}\left(k\bar{\h}\right)$ for sums of periodic perturbations
as well.

\bibliographystyle{apsrev}
\bibliography{JNP_abbr}

\begin{thebibliography}{62}
\expandafter\ifx\csname natexlab\endcsname\relax\def\natexlab#1{#1}\fi
\expandafter\ifx\csname bibnamefont\endcsname\relax
  \def\bibnamefont#1{#1}\fi
\expandafter\ifx\csname bibfnamefont\endcsname\relax
  \def\bibfnamefont#1{#1}\fi
\expandafter\ifx\csname citenamefont\endcsname\relax
  \def\citenamefont#1{#1}\fi
\expandafter\ifx\csname url\endcsname\relax
  \def\url#1{\texttt{#1}}\fi
\expandafter\ifx\csname urlprefix\endcsname\relax\def\urlprefix{URL }\fi
\providecommand{\bibinfo}[2]{#2}
\providecommand{\eprint}[2][]{\url{#2}}

\bibitem[{\citenamefont{Bimberg et~al.}(1999)\citenamefont{Bimberg, Grundmann,
  and Ledentsov}}]{Bimberg99}
\bibinfo{author}{\bibfnamefont{D.}~\bibnamefont{Bimberg}},
  \bibinfo{author}{\bibfnamefont{M.}~\bibnamefont{Grundmann}},
  \bibnamefont{and} \bibinfo{author}{\bibfnamefont{N.~N.}
  \bibnamefont{Ledentsov}}, \emph{\bibinfo{title}{Quantum Dot
  Heterostructures}} (\bibinfo{publisher}{John Wiley \& Sons},
  \bibinfo{address}{West Sussex, UK}, \bibinfo{year}{1999}).

\bibitem[{\citenamefont{Bayer et~al.}(2000)\citenamefont{Bayer, Stern,
  Hawrylak, Fafard, and Forchel}}]{Bayer2000}
\bibinfo{author}{\bibfnamefont{M.}~\bibnamefont{Bayer}},
  \bibinfo{author}{\bibfnamefont{O.}~\bibnamefont{Stern}},
  \bibinfo{author}{\bibfnamefont{P.}~\bibnamefont{Hawrylak}},
  \bibinfo{author}{\bibfnamefont{S.}~\bibnamefont{Fafard}}, \bibnamefont{and}
  \bibinfo{author}{\bibfnamefont{A.}~\bibnamefont{Forchel}},
  \bibinfo{journal}{Nature} \textbf{\bibinfo{volume}{405}},
  \bibinfo{pages}{923} (\bibinfo{year}{2000}).

\bibitem[{\citenamefont{Akiyama et~al.}(2000)\citenamefont{Akiyama, Kuwatsuka,
  Simoyama, Nakata, Mukai, Sugawara, Wada, and Ishikawa}}]{Akiyama:fiberoptics}
\bibinfo{author}{\bibfnamefont{T.}~\bibnamefont{Akiyama}},
  \bibinfo{author}{\bibfnamefont{H.}~\bibnamefont{Kuwatsuka}},
  \bibinfo{author}{\bibfnamefont{T.}~\bibnamefont{Simoyama}},
  \bibinfo{author}{\bibfnamefont{Y.}~\bibnamefont{Nakata}},
  \bibinfo{author}{\bibfnamefont{K.}~\bibnamefont{Mukai}},
  \bibinfo{author}{\bibfnamefont{M.}~\bibnamefont{Sugawara}},
  \bibinfo{author}{\bibfnamefont{O.}~\bibnamefont{Wada}}, \bibnamefont{and}
  \bibinfo{author}{\bibfnamefont{H.}~\bibnamefont{Ishikawa}},
  \bibinfo{journal}{IEEE Photonics Technology Letters}
  \textbf{\bibinfo{volume}{12}}, \bibinfo{pages}{1301} (\bibinfo{year}{2000}).

\bibitem[{\citenamefont{Viktorov and Mandel}(2006)}]{Viktorov:Laser}
\bibinfo{author}{\bibfnamefont{E.~A.} \bibnamefont{Viktorov}} \bibnamefont{and}
  \bibinfo{author}{\bibfnamefont{P.}~\bibnamefont{Mandel}},
  \bibinfo{journal}{Applied Physics Letters} \textbf{\bibinfo{volume}{88}},
  \bibinfo{pages}{201102} (\bibinfo{year}{2006}).

\bibitem[{\citenamefont{Kane}(1998)}]{Kane:QantumComputer}
\bibinfo{author}{\bibfnamefont{B.~E.} \bibnamefont{Kane}},
  \bibinfo{journal}{Nature} \textbf{\bibinfo{volume}{393}},
  \bibinfo{pages}{133} (\bibinfo{year}{1998}).

\bibitem[{\citenamefont{Elzerman et~al.}(2004)\citenamefont{Elzerman, Hanson,
  van Beveren, Witkamp, Vandersypen, and
  Kouwenhoven}}]{Elzerman:QuantumComputer}
\bibinfo{author}{\bibfnamefont{J.~M.} \bibnamefont{Elzerman}},
  \bibinfo{author}{\bibfnamefont{R.}~\bibnamefont{Hanson}},
  \bibinfo{author}{\bibfnamefont{L.~H.~W.} \bibnamefont{van Beveren}},
  \bibinfo{author}{\bibfnamefont{B.}~\bibnamefont{Witkamp}},
  \bibinfo{author}{\bibfnamefont{L.~M.~K.} \bibnamefont{Vandersypen}},
  \bibnamefont{and} \bibinfo{author}{\bibfnamefont{L.~P.}
  \bibnamefont{Kouwenhoven}}, \bibinfo{journal}{Nature}
  \textbf{\bibinfo{volume}{430}}, \bibinfo{pages}{431} (\bibinfo{year}{2004}).

\bibitem[{\citenamefont{Tanner et~al.}(2006)\citenamefont{Tanner, Hasko, and
  Williams}}]{Tanner:QuantumComputer}
\bibinfo{author}{\bibfnamefont{M.~G.} \bibnamefont{Tanner}},
  \bibinfo{author}{\bibfnamefont{D.~G.} \bibnamefont{Hasko}}, \bibnamefont{and}
  \bibinfo{author}{\bibfnamefont{D.~A.} \bibnamefont{Williams}},
  \bibinfo{journal}{Microelectronic Engineering} \textbf{\bibinfo{volume}{83}},
  \bibinfo{pages}{1818} (\bibinfo{year}{2006}).

\bibitem[{\citenamefont{Li et~al.}(1996)\citenamefont{Li, Xia, Yuan, Xu, Ge,
  Wang, Wang, Wang, and Chang}}]{Li:1996fk}
\bibinfo{author}{\bibfnamefont{S.-S.} \bibnamefont{Li}},
  \bibinfo{author}{\bibfnamefont{J.-B.} \bibnamefont{Xia}},
  \bibinfo{author}{\bibfnamefont{Z.~L.} \bibnamefont{Yuan}},
  \bibinfo{author}{\bibfnamefont{Z.~Y.} \bibnamefont{Xu}},
  \bibinfo{author}{\bibfnamefont{W.}~\bibnamefont{Ge}},
  \bibinfo{author}{\bibfnamefont{X.~R.} \bibnamefont{Wang}},
  \bibinfo{author}{\bibfnamefont{Y.}~\bibnamefont{Wang}},
  \bibinfo{author}{\bibfnamefont{J.}~\bibnamefont{Wang}}, \bibnamefont{and}
  \bibinfo{author}{\bibfnamefont{L.~L.} \bibnamefont{Chang}},
  \bibinfo{journal}{Phys. Rev. B} \textbf{\bibinfo{volume}{54}},
  \bibinfo{pages}{11575} (\bibinfo{year}{1996}).

\bibitem[{\citenamefont{Li and Xia}(1997)}]{Li:1997lr}
\bibinfo{author}{\bibfnamefont{S.-S.} \bibnamefont{Li}} \bibnamefont{and}
  \bibinfo{author}{\bibfnamefont{J.-B.} \bibnamefont{Xia}},
  \bibinfo{journal}{Phys. Rev. B} \textbf{\bibinfo{volume}{55}},
  \bibinfo{pages}{15434} (\bibinfo{year}{1997}).

\bibitem[{\citenamefont{Pchelyakov et~al.}(2000)\citenamefont{Pchelyakov,
  Bolkhovityanov, Dvurechenski, Sokolov, Nikiforov, Yakimov, and
  Voigtl{\"a}nder}}]{Pchelyakov2000}
\bibinfo{author}{\bibfnamefont{O.~P.} \bibnamefont{Pchelyakov}},
  \bibinfo{author}{\bibfnamefont{Y.~B.} \bibnamefont{Bolkhovityanov}},
  \bibinfo{author}{\bibfnamefont{A.~V.} \bibnamefont{Dvurechenski}},
  \bibinfo{author}{\bibfnamefont{L.~V.} \bibnamefont{Sokolov}},
  \bibinfo{author}{\bibfnamefont{A.~I.} \bibnamefont{Nikiforov}},
  \bibinfo{author}{\bibfnamefont{A.~I.} \bibnamefont{Yakimov}},
  \bibnamefont{and}
  \bibinfo{author}{\bibfnamefont{B.}~\bibnamefont{Voigtl{\"a}nder}},
  \bibinfo{journal}{Semiconductors} \textbf{\bibinfo{volume}{34}},
  \bibinfo{pages}{122947} (\bibinfo{year}{2000}),
  \bibinfo{note}{[\linkable{doi:}10.1134/1.1325416]}.

\bibitem[{\citenamefont{Grundmann}(2000)}]{Grundmann2000}
\bibinfo{author}{\bibfnamefont{M.}~\bibnamefont{Grundmann}},
  \bibinfo{journal}{Physica E} \textbf{\bibinfo{volume}{5}},
  \bibinfo{pages}{167} (\bibinfo{year}{2000}),
  \bibinfo{note}{[\linkable{doi:}10.1016/S1386-9477(99)00041-7]}.

\bibitem[{\citenamefont{Petroff et~al.}(2001)\citenamefont{Petroff, Lorke, and
  Imamoglu}}]{Petroff2001}
\bibinfo{author}{\bibfnamefont{P.}~\bibnamefont{Petroff}},
  \bibinfo{author}{\bibfnamefont{A.}~\bibnamefont{Lorke}}, \bibnamefont{and}
  \bibinfo{author}{\bibfnamefont{A.}~\bibnamefont{Imamoglu}},
  \bibinfo{journal}{Phys. Today} pp. \bibinfo{pages}{46--52}
  (\bibinfo{year}{2001}), \bibinfo{note}{[\linkable{doi:}10.1063/1.1381102]}.

\bibitem[{\citenamefont{Liu et~al.}(2001)\citenamefont{Liu, Xu, Wei, Ding,
  Qian, Han, Liang, and Wang}}]{Liu2001}
\bibinfo{author}{\bibfnamefont{H.-Y.} \bibnamefont{Liu}},
  \bibinfo{author}{\bibfnamefont{B.}~\bibnamefont{Xu}},
  \bibinfo{author}{\bibfnamefont{Y.-Q.} \bibnamefont{Wei}},
  \bibinfo{author}{\bibfnamefont{D.}~\bibnamefont{Ding}},
  \bibinfo{author}{\bibfnamefont{J.-J.} \bibnamefont{Qian}},
  \bibinfo{author}{\bibfnamefont{Q.}~\bibnamefont{Han}},
  \bibinfo{author}{\bibfnamefont{J.-B.} \bibnamefont{Liang}}, \bibnamefont{and}
  \bibinfo{author}{\bibfnamefont{Z.-G.} \bibnamefont{Wang}},
  \bibinfo{journal}{Appl. Phys. Lett.} \textbf{\bibinfo{volume}{79}},
  \bibinfo{pages}{2868} (\bibinfo{year}{2001}).

\bibitem[{\citenamefont{Heinrichsdorff
  et~al.}(1997)\citenamefont{Heinrichsdorff, Mao, Kirstaedter, Krost, Bimberg,
  Kosogov, and Werner}}]{Heinrichsdorff1997}
\bibinfo{author}{\bibfnamefont{F.}~\bibnamefont{Heinrichsdorff}},
  \bibinfo{author}{\bibfnamefont{M.}~\bibnamefont{Mao}},
  \bibinfo{author}{\bibfnamefont{N.}~\bibnamefont{Kirstaedter}},
  \bibinfo{author}{\bibfnamefont{A.}~\bibnamefont{Krost}},
  \bibinfo{author}{\bibfnamefont{D.}~\bibnamefont{Bimberg}},
  \bibinfo{author}{\bibfnamefont{A.}~\bibnamefont{Kosogov}}, \bibnamefont{and}
  \bibinfo{author}{\bibfnamefont{P.}~\bibnamefont{Werner}},
  \bibinfo{journal}{Appl. Phys. Lett.} \textbf{\bibinfo{volume}{71}},
  \bibinfo{pages}{22} (\bibinfo{year}{1997}),
  \bibinfo{note}{[\linkable{doi:}10.1063/1.120556]}.

\bibitem[{\citenamefont{Bimberg et~al.}(2002)\citenamefont{Bimberg, Ledentsov,
  and Lott}}]{Bimberg2002}
\bibinfo{author}{\bibfnamefont{D.}~\bibnamefont{Bimberg}},
  \bibinfo{author}{\bibfnamefont{N.}~\bibnamefont{Ledentsov}},
  \bibnamefont{and} \bibinfo{author}{\bibfnamefont{J.}~\bibnamefont{Lott}},
  \bibinfo{journal}{MRS Bull.} \textbf{\bibinfo{volume}{27}},
  \bibinfo{pages}{531} (\bibinfo{year}{2002}).

\bibitem[{\citenamefont{Friesen et~al.}(2003)\citenamefont{Friesen, Rugheimer,
  Savage, Lagally, van~der Weide, Joynt, and Eriksson}}]{Friesen2003}
\bibinfo{author}{\bibfnamefont{M.}~\bibnamefont{Friesen}},
  \bibinfo{author}{\bibfnamefont{P.}~\bibnamefont{Rugheimer}},
  \bibinfo{author}{\bibfnamefont{D.~E.} \bibnamefont{Savage}},
  \bibinfo{author}{\bibfnamefont{M.~G.} \bibnamefont{Lagally}},
  \bibinfo{author}{\bibfnamefont{D.~W.} \bibnamefont{van~der Weide}},
  \bibinfo{author}{\bibfnamefont{R.}~\bibnamefont{Joynt}}, \bibnamefont{and}
  \bibinfo{author}{\bibfnamefont{M.~A.} \bibnamefont{Eriksson}},
  \bibinfo{journal}{Phys. Rev. B} \textbf{\bibinfo{volume}{67}},
  \bibinfo{pages}{121301{ (R)}} (\bibinfo{year}{2003}),
  \bibinfo{note}{[\linkable{doi:}10.1103/PhysRevB.67.121301]}.

\bibitem[{\citenamefont{Cheng et~al.}(2003)\citenamefont{Cheng, Yang, Yang,
  Chang, and Hsieh}}]{Cheng2003}
\bibinfo{author}{\bibfnamefont{Y.-C.} \bibnamefont{Cheng}},
  \bibinfo{author}{\bibfnamefont{S.}~\bibnamefont{Yang}},
  \bibinfo{author}{\bibfnamefont{J.-N.} \bibnamefont{Yang}},
  \bibinfo{author}{\bibfnamefont{L.-B.} \bibnamefont{Chang}}, \bibnamefont{and}
  \bibinfo{author}{\bibfnamefont{L.-Z.} \bibnamefont{Hsieh}},
  \bibinfo{journal}{Opt. Eng.} \textbf{\bibinfo{volume}{42}},
  \bibinfo{pages}{11923} (\bibinfo{year}{2003}),
  \bibinfo{note}{[\linkable{doi:}doi:10.1117/1.1525277]}.

\bibitem[{\citenamefont{Krebs et~al.}(2003)\citenamefont{Krebs, Deubert,
  Reithmaier, and Forchel}}]{Krebs2003}
\bibinfo{author}{\bibfnamefont{R.}~\bibnamefont{Krebs}},
  \bibinfo{author}{\bibfnamefont{S.}~\bibnamefont{Deubert}},
  \bibinfo{author}{\bibfnamefont{J.}~\bibnamefont{Reithmaier}},
  \bibnamefont{and} \bibinfo{author}{\bibfnamefont{A.}~\bibnamefont{Forchel}},
  \bibinfo{journal}{J. Cryst. Growth} \textbf{\bibinfo{volume}{251}},
  \bibinfo{pages}{7427} (\bibinfo{year}{2003}),
  \bibinfo{note}{[\linkable{doi:}10.1016/S0022-0248(02)02385-0]}.

\bibitem[{\citenamefont{Sakaki}(2003)}]{Sakaki2003}
\bibinfo{author}{\bibfnamefont{H.}~\bibnamefont{Sakaki}}, \bibinfo{journal}{J.
  Cryst. Growth} \textbf{\bibinfo{volume}{251}}, \bibinfo{pages}{9}
  (\bibinfo{year}{2003}),
  \bibinfo{note}{[\linkable{doi:}10.1016/S0022-0248(03)00831-5]}.

\bibitem[{\citenamefont{Spencer et~al.}(1991)\citenamefont{Spencer, Voorhees,
  and Davis}}]{Spencer:1991we}
\bibinfo{author}{\bibfnamefont{B.~J.} \bibnamefont{Spencer}},
  \bibinfo{author}{\bibfnamefont{P.~W.} \bibnamefont{Voorhees}},
  \bibnamefont{and} \bibinfo{author}{\bibfnamefont{S.~H.} \bibnamefont{Davis}},
  \bibinfo{journal}{Phys. Rev. Lett.} \textbf{\bibinfo{volume}{67}},
  \bibinfo{pages}{3696} (\bibinfo{year}{1991}),
  \bibinfo{note}{[\linkable{doi:}10.1103/PhysRevLett.67.3696]}.

\bibitem[{\citenamefont{Spencer
  et~al.}(1993{\natexlab{a}})\citenamefont{Spencer, Voorhees, and
  Davis}}]{Spencer:1993vt}
\bibinfo{author}{\bibfnamefont{B.~J.} \bibnamefont{Spencer}},
  \bibinfo{author}{\bibfnamefont{P.~W.} \bibnamefont{Voorhees}},
  \bibnamefont{and} \bibinfo{author}{\bibfnamefont{S.~H.} \bibnamefont{Davis}},
  \bibinfo{journal}{J. Appl. Phys.} \textbf{\bibinfo{volume}{73}},
  \bibinfo{pages}{4955} (\bibinfo{year}{1993}{\natexlab{a}}),
  \bibinfo{note}{[\linkable{doi:} 10.1063/1.353815]}.

\bibitem[{\citenamefont{Tekalign and Spencer}(2004)}]{Tekalign:2004jh}
\bibinfo{author}{\bibfnamefont{W.~T.} \bibnamefont{Tekalign}} \bibnamefont{and}
  \bibinfo{author}{\bibfnamefont{B.~J.} \bibnamefont{Spencer}},
  \bibinfo{journal}{J. Appl. Phys.} \textbf{\bibinfo{volume}{96}},
  \bibinfo{pages}{5505} (\bibinfo{year}{2004}),
  \bibinfo{note}{[\linkable{doi:}10.1063/1.1766084]}.

\bibitem[{\citenamefont{Tekalign and Spencer}(2007)}]{Tekalign:2007}
\bibinfo{author}{\bibfnamefont{W.~T.} \bibnamefont{Tekalign}} \bibnamefont{and}
  \bibinfo{author}{\bibfnamefont{B.~J.} \bibnamefont{Spencer}},
  \bibinfo{journal}{J. Appl. Phys.} \textbf{\bibinfo{volume}{102}},
  \bibinfo{pages}{073503} (\bibinfo{year}{2007}).

\bibitem[{\citenamefont{Obayashi and Shintani}(1998)}]{Obayashi:1998fk}
\bibinfo{author}{\bibfnamefont{Y.}~\bibnamefont{Obayashi}} \bibnamefont{and}
  \bibinfo{author}{\bibfnamefont{K.}~\bibnamefont{Shintani}},
  \bibinfo{journal}{J. Appl. Phys.} \textbf{\bibinfo{volume}{84}},
  \bibinfo{pages}{3141} (\bibinfo{year}{1998}),
  \bibinfo{note}{[\linkable{doi:}10.1063/1.368468]}.

\bibitem[{\citenamefont{Ross et~al.}(1998)\citenamefont{Ross, Tersoff, and
  Tromp}}]{Ross:1998fk}
\bibinfo{author}{\bibfnamefont{F.~M.} \bibnamefont{Ross}},
  \bibinfo{author}{\bibfnamefont{J.}~\bibnamefont{Tersoff}}, \bibnamefont{and}
  \bibinfo{author}{\bibfnamefont{R.~M.} \bibnamefont{Tromp}},
  \bibinfo{journal}{Phys. Rev. Lett.} \textbf{\bibinfo{volume}{80}},
  \bibinfo{pages}{984} (\bibinfo{year}{1998}),
  \bibinfo{note}{[\linkable{doi:}10.1103/PhysRevLett.80.984]}.

\bibitem[{\citenamefont{Ozkan et~al.}(1999)\citenamefont{Ozkan, Nix, and
  Gao}}]{Ozkan:1999gf}
\bibinfo{author}{\bibfnamefont{C.~S.} \bibnamefont{Ozkan}},
  \bibinfo{author}{\bibfnamefont{W.~D.} \bibnamefont{Nix}}, \bibnamefont{and}
  \bibinfo{author}{\bibfnamefont{H.~J.} \bibnamefont{Gao}},
  \bibinfo{journal}{J. Mater. Res.} \textbf{\bibinfo{volume}{14}},
  \bibinfo{pages}{3247} (\bibinfo{year}{1999}),
  \bibinfo{note}{[\linkable{doi:}10.1557/JMR.1999.043]}.

\bibitem[{\citenamefont{Gao and Nix}(1999)}]{Gao:1999ve}
\bibinfo{author}{\bibfnamefont{H.~J.} \bibnamefont{Gao}} \bibnamefont{and}
  \bibinfo{author}{\bibfnamefont{W.~D.} \bibnamefont{Nix}},
  \bibinfo{journal}{Ann. Rev. Mater. Sci.} \textbf{\bibinfo{volume}{29}},
  \bibinfo{pages}{173} (\bibinfo{year}{1999}),
  \bibinfo{note}{[\linkable{doi:}0.1146/annurev.matsci.29.1.173]}.

\bibitem[{\citenamefont{Holy et~al.}(1999)\citenamefont{Holy, Springholz,
  Pinczolits, and Bauer}}]{Holy:1999th}
\bibinfo{author}{\bibfnamefont{V.}~\bibnamefont{Holy}},
  \bibinfo{author}{\bibfnamefont{G.}~\bibnamefont{Springholz}},
  \bibinfo{author}{\bibfnamefont{M.}~\bibnamefont{Pinczolits}},
  \bibnamefont{and} \bibinfo{author}{\bibfnamefont{G.}~\bibnamefont{Bauer}},
  \bibinfo{journal}{Phys. Rev. Lett.} \textbf{\bibinfo{volume}{83}},
  \bibinfo{pages}{356} (\bibinfo{year}{1999}),
  \bibinfo{note}{[\linkable{doi:}10.1103/PhysRevLett.83.356]}.

\bibitem[{\citenamefont{Ortiz et~al.}(1999)\citenamefont{Ortiz, Repetto, and
  Si}}]{Ortiz:1999ys}
\bibinfo{author}{\bibfnamefont{M.}~\bibnamefont{Ortiz}},
  \bibinfo{author}{\bibfnamefont{E.}~\bibnamefont{Repetto}}, \bibnamefont{and}
  \bibinfo{author}{\bibfnamefont{H.}~\bibnamefont{Si}},
  \bibinfo{journal}{Journal of the Mechanics and Physics of Solids}
  \textbf{\bibinfo{volume}{47}}, \bibinfo{pages}{697} (\bibinfo{year}{1999}).

\bibitem[{\citenamefont{Zhang et~al.}(2003)\citenamefont{Zhang, Bower, and
  Liu}}]{Zhang:2003tg}
\bibinfo{author}{\bibfnamefont{Y.}~\bibnamefont{Zhang}},
  \bibinfo{author}{\bibfnamefont{A.}~\bibnamefont{Bower}}, \bibnamefont{and}
  \bibinfo{author}{\bibfnamefont{P.}~\bibnamefont{Liu}}, \bibinfo{journal}{Thin
  Solid Films} \textbf{\bibinfo{volume}{424}}, \bibinfo{pages}{9}
  (\bibinfo{year}{2003}),
  \bibinfo{note}{[\linkable{doi:}10.1016/S0040-6090(02)00897-0]}.

\bibitem[{\citenamefont{Zhang and Bower}(2001)}]{Zhang:2001cr}
\bibinfo{author}{\bibfnamefont{Y.~W.} \bibnamefont{Zhang}} \bibnamefont{and}
  \bibinfo{author}{\bibfnamefont{A.~F.} \bibnamefont{Bower}},
  \bibinfo{journal}{Appl. Phys. Lett.} \textbf{\bibinfo{volume}{78}},
  \bibinfo{pages}{2706} (\bibinfo{year}{2001}),
  \bibinfo{note}{[\linkable{doi:}10.1063/1.1354155]}.

\bibitem[{\citenamefont{Meixner et~al.}(2003)\citenamefont{Meixner, Kunert, and
  Scholl}}]{Meixner:2003lz}
\bibinfo{author}{\bibfnamefont{M.}~\bibnamefont{Meixner}},
  \bibinfo{author}{\bibfnamefont{R.}~\bibnamefont{Kunert}}, \bibnamefont{and}
  \bibinfo{author}{\bibfnamefont{E.}~\bibnamefont{Scholl}},
  \bibinfo{journal}{Phys. Rev. B} \textbf{\bibinfo{volume}{67}},
  \bibinfo{pages}{195301} (\bibinfo{year}{2003}),
  \bibinfo{note}{[\linkable{doi:} 10.1103/PhysRevB.67.195301]}.

\bibitem[{\citenamefont{Liu et~al.}(2003{\natexlab{a}})\citenamefont{Liu,
  Zhang, and Lu}}]{Liu:2003kx}
\bibinfo{author}{\bibfnamefont{P.}~\bibnamefont{Liu}},
  \bibinfo{author}{\bibfnamefont{Y.~W.} \bibnamefont{Zhang}}, \bibnamefont{and}
  \bibinfo{author}{\bibfnamefont{C.}~\bibnamefont{Lu}}, \bibinfo{journal}{Phys.
  Rev. B} \textbf{\bibinfo{volume}{67}}, \bibinfo{pages}{165414}
  (\bibinfo{year}{2003}{\natexlab{a}}), \bibinfo{note}{[\linkable{doi:}
  10.1103/PhysRevB.67.165414]}.

\bibitem[{\citenamefont{Liu et~al.}(2003{\natexlab{b}})\citenamefont{Liu,
  Zhang, and Lu}}]{Liu:2003qi}
\bibinfo{author}{\bibfnamefont{P.}~\bibnamefont{Liu}},
  \bibinfo{author}{\bibfnamefont{Y.~W.} \bibnamefont{Zhang}}, \bibnamefont{and}
  \bibinfo{author}{\bibfnamefont{C.}~\bibnamefont{Lu}}, \bibinfo{journal}{Phys.
  Rev. B} \textbf{\bibinfo{volume}{68}}, \bibinfo{pages}{035402}
  (\bibinfo{year}{2003}{\natexlab{b}}),
  \bibinfo{note}{[\linkable{doi:}10.1103/PhysRevB.68.035402]}.

\bibitem[{\citenamefont{Golovin et~al.}(2003)\citenamefont{Golovin, Davis, and
  Voorhees}}]{Golovin:2003ms}
\bibinfo{author}{\bibfnamefont{A.~A.} \bibnamefont{Golovin}},
  \bibinfo{author}{\bibfnamefont{S.~H.} \bibnamefont{Davis}}, \bibnamefont{and}
  \bibinfo{author}{\bibfnamefont{P.~W.} \bibnamefont{Voorhees}},
  \bibinfo{journal}{Phys. Rev. E} \textbf{\bibinfo{volume}{68}},
  \bibinfo{pages}{056203} (\bibinfo{year}{2003}),
  \bibinfo{note}{[\linkable{doi:}10.1103/PhysRevE.68.056203]}.

\bibitem[{\citenamefont{Friedman}(2007{\natexlab{a}})}]{FriedmanStochastic}
\bibinfo{author}{\bibfnamefont{L.~H.} \bibnamefont{Friedman}},
  \bibinfo{journal}{J. of Electron. Mater.} \textbf{\bibinfo{volume}{36}},
  \bibinfo{pages}{1546} (\bibinfo{year}{2007}{\natexlab{a}}).

\bibitem[{\citenamefont{Friedman}(2007{\natexlab{b}})}]{FriedmanJNP07}
\bibinfo{author}{\bibfnamefont{L.~H.} \bibnamefont{Friedman}},
  \bibinfo{journal}{J. of Nanophotonics} \textbf{\bibinfo{volume}{1}},
  \bibinfo{pages}{013513} (\bibinfo{year}{2007}{\natexlab{b}}).

\bibitem[{\citenamefont{Kumar and Friedman}(2007)}]{Kumar:2007fk}
\bibinfo{author}{\bibfnamefont{C.}~\bibnamefont{Kumar}} \bibnamefont{and}
  \bibinfo{author}{\bibfnamefont{L.~H.} \bibnamefont{Friedman}},
  \bibinfo{journal}{J. Appl. Phys.} \textbf{\bibinfo{volume}{101}},
  \bibinfo{eid}{094903} (\bibinfo{year}{2007}).

\bibitem[{\citenamefont{Ramasubramaniam and
  Shenoy}(2005{\natexlab{a}})}]{Ramasubramaniam:2005oq}
\bibinfo{author}{\bibfnamefont{A.}~\bibnamefont{Ramasubramaniam}}
  \bibnamefont{and} \bibinfo{author}{\bibfnamefont{V.~B.}
  \bibnamefont{Shenoy}}, \bibinfo{journal}{J. Eng. Mater.-T. ASME}
  \textbf{\bibinfo{volume}{127}}, \bibinfo{pages}{434}
  (\bibinfo{year}{2005}{\natexlab{a}}),
  \bibinfo{note}{[\linkable{doi:}10.1115/1.1924559]}.

\bibitem[{\citenamefont{Ramasubramaniam and
  Shenoy}(2005{\natexlab{b}})}]{Ramasubramaniam:2005vn}
\bibinfo{author}{\bibfnamefont{A.}~\bibnamefont{Ramasubramaniam}}
  \bibnamefont{and} \bibinfo{author}{\bibfnamefont{V.~B.}
  \bibnamefont{Shenoy}}, \bibinfo{journal}{J. Appl. Phys.}
  \textbf{\bibinfo{volume}{97}}, \bibinfo{eid}{114312}
  (\bibinfo{year}{2005}{\natexlab{b}}), \bibinfo{note}{[\linkable{doi:}
  10.1063/1.1897837]}.

\bibitem[{\citenamefont{Niu et~al.}(2006)\citenamefont{Niu, Vardavas, Caflisch,
  and Ratsch}}]{Niu:2006fk}
\bibinfo{author}{\bibfnamefont{X.}~\bibnamefont{Niu}},
  \bibinfo{author}{\bibfnamefont{R.}~\bibnamefont{Vardavas}},
  \bibinfo{author}{\bibfnamefont{R.~E.} \bibnamefont{Caflisch}},
  \bibnamefont{and} \bibinfo{author}{\bibfnamefont{C.}~\bibnamefont{Ratsch}},
  \bibinfo{journal}{Phys. Rev. B} \textbf{\bibinfo{volume}{74}},
  \bibinfo{pages}{193403} (\bibinfo{year}{2006}), ISSN
  \bibinfo{issn}{1098-0121},
  \bibinfo{note}{[\linkable{doi:}10.1103/PhysRevB.74.193403}.

\bibitem[{\citenamefont{Spencer
  et~al.}(1993{\natexlab{b}})\citenamefont{Spencer, Davis, and
  Voorhees}}]{Spencer:1993ve}
\bibinfo{author}{\bibfnamefont{B.~J.} \bibnamefont{Spencer}},
  \bibinfo{author}{\bibfnamefont{S.~H.} \bibnamefont{Davis}}, \bibnamefont{and}
  \bibinfo{author}{\bibfnamefont{P.~W.} \bibnamefont{Voorhees}},
  \bibinfo{journal}{Phys. Rev. Lett.} \textbf{\bibinfo{volume}{47}},
  \bibinfo{pages}{9760} (\bibinfo{year}{1993}{\natexlab{b}}),
  \bibinfo{note}{[\linkable{doi:} 10.1103/PhysRevB.47.9760]}.

\bibitem[{\citenamefont{Friedman}(2007{\natexlab{c}})}]{Friedman:fk}
\bibinfo{author}{\bibfnamefont{L.~H.} \bibnamefont{Friedman}},
  \bibinfo{journal}{Phys. Rev. B} \textbf{\bibinfo{volume}{75}},
  \bibinfo{pages}{193302} (\bibinfo{year}{2007}{\natexlab{c}}).

\bibitem[{\citenamefont{Asaro and Tiller}(1972)}]{Asaro-Tiller}
\bibinfo{author}{\bibfnamefont{R.~J.} \bibnamefont{Asaro}} \bibnamefont{and}
  \bibinfo{author}{\bibfnamefont{W.~A.} \bibnamefont{Tiller}},
  \bibinfo{journal}{Met. Trans.} \textbf{\bibinfo{volume}{3}},
  \bibinfo{pages}{1789} (\bibinfo{year}{1972}).

\bibitem[{\citenamefont{Grinfeld}(1986)}]{Grinfeld}
\bibinfo{author}{\bibfnamefont{M.~A.} \bibnamefont{Grinfeld}},
  \bibinfo{journal}{Sov. Phys. Dokl.} \textbf{\bibinfo{volume}{31}},
  \bibinfo{pages}{831} (\bibinfo{year}{1986}).

\bibitem[{\citenamefont{Wang et~al.}(2004)\citenamefont{Wang, Jin, and
  Khachaturyan}}]{Wang:2004dd}
\bibinfo{author}{\bibfnamefont{Y.~U.} \bibnamefont{Wang}},
  \bibinfo{author}{\bibfnamefont{Y.~M.} \bibnamefont{Jin}}, \bibnamefont{and}
  \bibinfo{author}{\bibfnamefont{A.~G.} \bibnamefont{Khachaturyan}},
  \bibinfo{journal}{Acta Mater.} \textbf{\bibinfo{volume}{52}},
  \bibinfo{pages}{81} (\bibinfo{year}{2004}),
  \bibinfo{note}{[\linkable{doi:}10.1016/j.actamat.2003.08.027]}.

\bibitem[{\citenamefont{Rastelli et~al.}(2005)\citenamefont{Rastelli, Stoffel,
  Tersoff, Kar, and Schmidt}}]{Rastelli:2005kl}
\bibinfo{author}{\bibfnamefont{A.}~\bibnamefont{Rastelli}},
  \bibinfo{author}{\bibfnamefont{M.}~\bibnamefont{Stoffel}},
  \bibinfo{author}{\bibfnamefont{J.}~\bibnamefont{Tersoff}},
  \bibinfo{author}{\bibfnamefont{G.~S.} \bibnamefont{Kar}}, \bibnamefont{and}
  \bibinfo{author}{\bibfnamefont{O.~G.} \bibnamefont{Schmidt}},
  \bibinfo{journal}{Phys. Rev. Lett.} \textbf{\bibinfo{volume}{95}},
  \bibinfo{pages}{026103} (\bibinfo{year}{2005}), ISSN
  \bibinfo{issn}{0031-9007}.

\bibitem[{\citenamefont{Rastelli et~al.}(2006)\citenamefont{Rastelli, Stoffel,
  Denker, Merdzhanova, and Schmidt}}]{Rastelli:2006wd}
\bibinfo{author}{\bibfnamefont{A.}~\bibnamefont{Rastelli}},
  \bibinfo{author}{\bibfnamefont{M.}~\bibnamefont{Stoffel}},
  \bibinfo{author}{\bibfnamefont{U.}~\bibnamefont{Denker}},
  \bibinfo{author}{\bibfnamefont{T.}~\bibnamefont{Merdzhanova}},
  \bibnamefont{and} \bibinfo{author}{\bibfnamefont{O.~G.}
  \bibnamefont{Schmidt}}, \bibinfo{journal}{physica status solidi (a)}
  \textbf{\bibinfo{volume}{203}}, \bibinfo{pages}{3506} (\bibinfo{year}{2006}).

\bibitem[{\citenamefont{Tu and Tersoff}(2004)}]{Tu:2004tg}
\bibinfo{author}{\bibfnamefont{Y.~H.} \bibnamefont{Tu}} \bibnamefont{and}
  \bibinfo{author}{\bibfnamefont{J.}~\bibnamefont{Tersoff}},
  \bibinfo{journal}{Phys. Rev. Lett.} \textbf{\bibinfo{volume}{93}},
  \bibinfo{pages}{216101} (\bibinfo{year}{2004}), ISSN
  \bibinfo{issn}{0031-9007},
  \bibinfo{note}{[\linkable{doi:}10.1103/PhysRevLett.93.216101}.

\bibitem[{\citenamefont{Suo and Zhang}(1998)}]{suo98}
\bibinfo{author}{\bibfnamefont{Z.}~\bibnamefont{Suo}} \bibnamefont{and}
  \bibinfo{author}{\bibfnamefont{Z.}~\bibnamefont{Zhang}},
  \bibinfo{journal}{Phys. Rev. B} \textbf{\bibinfo{volume}{58}},
  \bibinfo{pages}{5116} (\bibinfo{year}{1998}).

\bibitem[{\citenamefont{Beck et~al.}(2004)\citenamefont{Beck, van~de Walle, and
  Asta}}]{Beck:2004yq}
\bibinfo{author}{\bibfnamefont{M.~J.} \bibnamefont{Beck}},
  \bibinfo{author}{\bibfnamefont{A.}~\bibnamefont{van~de Walle}},
  \bibnamefont{and} \bibinfo{author}{\bibfnamefont{M.}~\bibnamefont{Asta}},
  \bibinfo{journal}{Phys. Rev. B} \textbf{\bibinfo{volume}{70}},
  \bibinfo{pages}{205337} (\bibinfo{year}{2004}),
  \bibinfo{note}{[\linkable{doi:}10.1103/PhysRevB.70.205337]}.

\bibitem[{\citenamefont{Shenoy and Freund}(2002)}]{Shenoy:2002lr}
\bibinfo{author}{\bibfnamefont{V.~B.} \bibnamefont{Shenoy}} \bibnamefont{and}
  \bibinfo{author}{\bibfnamefont{L.~B.} \bibnamefont{Freund}},
  \bibinfo{journal}{Journal of the Mechanics and Physics of Solids}
  \textbf{\bibinfo{volume}{50}}, \bibinfo{pages}{1817} (\bibinfo{year}{2002}).

\bibitem[{\citenamefont{Tersoff et~al.}(2002)\citenamefont{Tersoff, Spencer,
  Rastelli, and von K\"anel}}]{Tersoff:2002qf}
\bibinfo{author}{\bibfnamefont{J.}~\bibnamefont{Tersoff}},
  \bibinfo{author}{\bibfnamefont{B.~J.} \bibnamefont{Spencer}},
  \bibinfo{author}{\bibfnamefont{A.}~\bibnamefont{Rastelli}}, \bibnamefont{and}
  \bibinfo{author}{\bibfnamefont{H.}~\bibnamefont{von K\"anel}},
  \bibinfo{journal}{Phys. Rev. Lett.} \textbf{\bibinfo{volume}{89}},
  \bibinfo{pages}{196104} (\bibinfo{year}{2002}).

\bibitem[{\citenamefont{Hong et~al.}(2006)\citenamefont{Hong, Kim, Lee, kwack,
  Han, and Oh}}]{Hong2006}
\bibinfo{author}{\bibfnamefont{S.~U.} \bibnamefont{Hong}},
  \bibinfo{author}{\bibfnamefont{J.~S.} \bibnamefont{Kim}},
  \bibinfo{author}{\bibfnamefont{J.~H.} \bibnamefont{Lee}},
  \bibinfo{author}{\bibfnamefont{H.-S.} \bibnamefont{kwack}},
  \bibinfo{author}{\bibfnamefont{W.-S.} \bibnamefont{Han}}, \bibnamefont{and}
  \bibinfo{author}{\bibfnamefont{D.~K.} \bibnamefont{Oh}}, \bibinfo{journal}{J.
  Cryst. Growth} \textbf{\bibinfo{volume}{286}}, \bibinfo{pages}{18}
  (\bibinfo{year}{2006}).

\bibitem[{\citenamefont{Zhao et~al.}(2006)\citenamefont{Zhao, Yoon, Feng, Li,
  Kim, Liu, Hulko, Xie, Kim, Kim et~al.}}]{Zhao:2006qy}
\bibinfo{author}{\bibfnamefont{Z.~M.} \bibnamefont{Zhao}},
  \bibinfo{author}{\bibfnamefont{T.~S.} \bibnamefont{Yoon}},
  \bibinfo{author}{\bibfnamefont{W.}~\bibnamefont{Feng}},
  \bibinfo{author}{\bibfnamefont{B.~Y.} \bibnamefont{Li}},
  \bibinfo{author}{\bibfnamefont{J.~H.} \bibnamefont{Kim}},
  \bibinfo{author}{\bibfnamefont{J.}~\bibnamefont{Liu}},
  \bibinfo{author}{\bibfnamefont{O.}~\bibnamefont{Hulko}},
  \bibinfo{author}{\bibfnamefont{Y.~H.} \bibnamefont{Xie}},
  \bibinfo{author}{\bibfnamefont{H.~M.} \bibnamefont{Kim}},
  \bibinfo{author}{\bibfnamefont{K.~B.} \bibnamefont{Kim}},
  \bibnamefont{et~al.}, \bibinfo{journal}{Thin Solid Films}
  \textbf{\bibinfo{volume}{508}}, \bibinfo{pages}{195} (\bibinfo{year}{2006}),
  ISSN \bibinfo{issn}{0040-6090},
  \bibinfo{note}{[\linkable{doi:}10.1016/j.tsf.2005.08.407]}.

\bibitem[{\citenamefont{Liu et~al.}(2003{\natexlab{c}})\citenamefont{Liu,
  Zhang, and Lu}}]{Liu:2003ig}
\bibinfo{author}{\bibfnamefont{P.}~\bibnamefont{Liu}},
  \bibinfo{author}{\bibfnamefont{Y.~W.} \bibnamefont{Zhang}}, \bibnamefont{and}
  \bibinfo{author}{\bibfnamefont{C.}~\bibnamefont{Lu}}, \bibinfo{journal}{Phys.
  Rev. B} \textbf{\bibinfo{volume}{68}}, \bibinfo{pages}{195314}
  (\bibinfo{year}{2003}{\natexlab{c}}),
  \bibinfo{note}{[\linkable{doi:}10.1103/PhysRevB.68.195314]}.

\bibitem[{\citenamefont{Golovin et~al.}(2004)\citenamefont{Golovin, Levine,
  Savina, and Davis}}]{Golovin:2004sa}
\bibinfo{author}{\bibfnamefont{A.~A.} \bibnamefont{Golovin}},
  \bibinfo{author}{\bibfnamefont{M.~S.} \bibnamefont{Levine}},
  \bibinfo{author}{\bibfnamefont{T.~V.} \bibnamefont{Savina}},
  \bibnamefont{and} \bibinfo{author}{\bibfnamefont{S.~H.} \bibnamefont{Davis}},
  \bibinfo{journal}{Phys. Rev. B} \textbf{\bibinfo{volume}{70}},
  \bibinfo{pages}{235342} (\bibinfo{year}{2004}),
  \bibinfo{note}{[\linkable{doi:}10.1103/PhysRevB.70.235342]}.

\bibitem[{\citenamefont{Freund and Suresh}(2003)}]{Freund:2003ih}
\bibinfo{author}{\bibfnamefont{L.~B.} \bibnamefont{Freund}} \bibnamefont{and}
  \bibinfo{author}{\bibfnamefont{S.}~\bibnamefont{Suresh}},
  \emph{\bibinfo{title}{Thin Film Materials: Stress, Defect Formation and
  Surface Evolution}} (\bibinfo{publisher}{Cambridge University Press},
  \bibinfo{address}{Cambridge, UK}, \bibinfo{year}{2003}),
  chap.~\bibinfo{chapter}{8}.

\bibitem[{\citenamefont{Vorbyev}(1996)}]{Vorbyev:1996fk}
\bibinfo{author}{\bibfnamefont{L.~E.} \bibnamefont{Vorbyev}},
  \emph{\bibinfo{title}{Handbook Series On Semiconductor Parameters}}
  (\bibinfo{publisher}{World Scientific}, \bibinfo{address}{Singapore},
  \bibinfo{year}{1996}), vol.~\bibinfo{volume}{1}.

\bibitem[{\citenamefont{Hegazy and Elsayed-Ali}(2006)}]{Hegazy:2006}
\bibinfo{author}{\bibfnamefont{M.~S.} \bibnamefont{Hegazy}} \bibnamefont{and}
  \bibinfo{author}{\bibfnamefont{H.~E.} \bibnamefont{Elsayed-Ali}},
  \bibinfo{journal}{J. Appl. Phys.} \textbf{\bibinfo{volume}{99}},
  \bibinfo{pages}{054308} (\bibinfo{year}{2006}).

\bibitem[{\citenamefont{Stangl et~al.}(2000)\citenamefont{Stangl, Roch, Holy,
  Pinczolits, Springholz, Bauer, Kegel, Metzger, Zhu, Brunner
  et~al.}}]{Stangl:2000superlattice}
\bibinfo{author}{\bibfnamefont{J.}~\bibnamefont{Stangl}},
  \bibinfo{author}{\bibfnamefont{T.}~\bibnamefont{Roch}},
  \bibinfo{author}{\bibfnamefont{V.}~\bibnamefont{Holy}},
  \bibinfo{author}{\bibfnamefont{M.}~\bibnamefont{Pinczolits}},
  \bibinfo{author}{\bibfnamefont{G.}~\bibnamefont{Springholz}},
  \bibinfo{author}{\bibfnamefont{G.}~\bibnamefont{Bauer}},
  \bibinfo{author}{\bibfnamefont{I.}~\bibnamefont{Kegel}},
  \bibinfo{author}{\bibfnamefont{T.~H.} \bibnamefont{Metzger}},
  \bibinfo{author}{\bibfnamefont{J.}~\bibnamefont{Zhu}},
  \bibinfo{author}{\bibfnamefont{K.}~\bibnamefont{Brunner}},
  \bibnamefont{et~al.}, \bibinfo{journal}{J. Vac. Sci. Technol. B}
  \textbf{\bibinfo{volume}{18}} (\bibinfo{year}{2000}).

\bibitem[{\citenamefont{Springholz et~al.}(2001)\citenamefont{Springholz,
  Pinczolits, Holy, Zerlauth, Vavra, and Bauer}}]{Springholz:2001nx}
\bibinfo{author}{\bibfnamefont{G.}~\bibnamefont{Springholz}},
  \bibinfo{author}{\bibfnamefont{M.}~\bibnamefont{Pinczolits}},
  \bibinfo{author}{\bibfnamefont{V.}~\bibnamefont{Holy}},
  \bibinfo{author}{\bibfnamefont{S.}~\bibnamefont{Zerlauth}},
  \bibinfo{author}{\bibfnamefont{I.}~\bibnamefont{Vavra}}, \bibnamefont{and}
  \bibinfo{author}{\bibfnamefont{G.}~\bibnamefont{Bauer}},
  \bibinfo{journal}{Physica E} \textbf{\bibinfo{volume}{9}},
  \bibinfo{pages}{149} (\bibinfo{year}{2001}),
  \bibinfo{note}{[\linkable{doi:}10.1016/S1386-9477(00)00189-2}.

\end{thebibliography}

\end{document}